\documentclass[preprint]{aastex}
\newcommand{\ecs}{erg cm$^{-2}$ s$^{-1}$}
\newcommand{\es}{erg s$^{-1}$}

\begin{document}

\title{\textit{Chandra} Observations of Candidate ``True'' Seyfert 2 Nuclei}
\author{Himel Ghosh, Richard W. Pogge, Smita Mathur, Paul Martini}
\affil{Department of Astronomy, The Ohio State University\\ 140 W 18th
Ave, Columbus, OH 43210}
\email{ghosh,pogge,smita,martini@astronomy.ohio-state.edu}
\and
\author{Joseph C. Shields}
\affil{Department of Physics \& Astronomy, Ohio University\\ Clippinger Lab 251B, Athens, OH 45701}
\email{shields@phy.ohiou.edu}

\begin{abstract}

The Unification Model for active galactic nuclei posits that Seyfert 2s
are intrinsically like Seyfert 1s, but that their broad-line regions
(BLRs) are hidden from our view. A Seyfert 2 nucleus that truly lacked a
BLR, instead of simply having it hidden, would be a so-called ``true''
Seyfert 2. No object has as yet been conclusively proven to be one. We
present a detailed analysis of four of the best ``true'' Seyfert 2
candidates discovered to date: IC 3639, NGC 3982, NGC 5283, and NGC
5427. None of the four has a broad H$\alpha$ emission line, either in
direct or polarized light. All four have rich, high-excitation spectra,
blue continua, and \textit{Hubble Space Telescope} (\textit{HST}) images
showing them to be unresolved sources with no host-galaxy
obscuration. To check for possible obscuration on scales smaller than
that resolvable by \textit{HST}, we obtained X-ray observations using
the \textit{Chandra X-ray Observatory}. All four objects show evidence
of obscuration and therefore could have hidden BLRs. The picture that
emerges is of moderate to high, but not necessarily Compton-thick,
obscuration of the nucleus, with extra-nuclear soft emission extended on
the hundreds-of-parsecs scale that may originate in the narrow-line
region. Since the extended soft emission compensates, in part, for the
nuclear soft emission lost to absorption, both absorption and luminosity
are likely to be severely underestimated unless the X-ray spectrum is of
sufficient quality to distinguish the two components. This is of special
concern where the source is too faint to produce a large number of
counts, or where the source is too far away to resolve the extended soft
X-ray emitting region.

\end{abstract}


\section{Introduction}

Seyfert Galaxies are relatively low-luminosity Active Galactic Nuclei
(AGNs) that are traditionally divided into two types, 1 and 2, following
an empirical optical spectroscopic classification scheme proposed over
30 years ago by \citet{kh74}.  Seyfert 1s are characterized by strong,
broad (widths of several thousand km\,s$^{-1}$) permitted emission lines
of Hydrogen, Helium, and \ion{Fe}{2} and narrow (typical widths of
$500-1000$\,km\,s$^{-1}$) forbidden lines such as [\ion{O}{3}],
[\ion{N}{2}], and [\ion{S}{2}]. Seyfert 2s have only
narrow permitted and forbidden lines of comparable width and are characterized
by a very high [\ion{O}{3}]$\lambda5007$/H$\beta$ line ratio.  Both
classes also often exhibit a blue featureless continuum, often
manifested as an ``ultraviolet excess,'' though this is not a necessary
criterion.

The leading explanation for the spectral differences between the two
classes of Seyferts is the Unification Model originally motivated by
observations of polarized broad emission lines in a number of Seyfert 2
galaxies \citep[and references therein]{ski93}.  Unification posits that
all Seyfert galaxies have the same underlying structure: a central
continuum source arising from an accretion disk around a central
supermassive black hole, a dense broad-line region (BLR) comprised of
high-velocity gas located on light-day scales, and a low-density
narrow-line region (NLR) on parsec scales.  The BLR is nestled inside a
torus of dusty obscuring material surrounding the nucleus, and whether
we see a Type 1 or 2 spectrum depends on the orientation of this torus
with respect to our line of sight \citep[see][Fig 7.1]{peterson97}.  In
Seyfert 1s we are looking down the opening of the torus and see the BLR
and central continuum source, while in Seyfert 2s the torus blocks the
central regions from view and we only see the narrow-line spectrum of
the NLR.  In some Seyfert 2s a circumnuclear cloud of gas in a favorable
location outside the torus opening acts as a polarizing ``mirror'' that
affords an oblique view down the unobscured axis of the torus, and we
see the BLR and underlying continuum in polarized light.  A large
fraction of the Seyfert 2s observed with spectropolarimetry to date have
polarized broad lines \citep{moran00,tran01,l01,lah04}.

Additional support for the Unification scenario comes from X-ray studies
that have demonstrated that many Seyfert 2s are highly absorbed at soft
X-rays by very large column densities 
\citep[$\log N_H> 23$, e.g.,][]{mea94,maiolino98,g01}.  Indeed, this idea
has become so compelling that many obscured AGNs, especially those that
are X-ray bright but have no or only very weak optical counterparts, are
called ``Type 2 AGN'', often regardless of whether or not their optical
spectra satisfy the original \citeauthor*{kh74} criteria.

While there is little doubt that a large fraction of Seyfert 2s are
indeed obscured Seyfert 1s, there is still no good answer to the
question of whether or not there exists a subset of
spectroscopically-classified Seyfert 2s that intrinsically lack a BLR.
Such BLR-free Seyfert 2s have variously been called ``true Seyfert 2s''
\citep[e.g.,][]{bd86} or ``pure Seyfert 2s'' \citep[e.g.,][]{h95} to
distinguish them from the Seyfert 2s that are hidden Seyfert 1s in the
Unification picture.  We shall adopt the term ``true Seyfert 2s'' in
this investigation.  Our motivation in taking up this question is not to
attempt to disprove the Unification picture, but rather to try to find
the interesting exceptions to it.  Identifying and studying objects that
lie at the extremes of the parameter space for the accretion process are
critical for informing us about the important physical details of that
process.

Why might there be no BLR in some AGN?  Recent observations and
theoretical work have suggested that the gas responsible for the BLR is
related, at least in part, to the accretion process
\citep[e.g., a high-velocity outflow from the disk,][]{mc98}.  One
possibility is that as an AGN becomes active, there is a brief period
where conditions conducive to formation of a BLR have not yet been
established.  Similarly, when an AGN turns off there might come a time
during the shutdown when the BLR can no longer be sustained.  The high
densities and small scale (few light days) of the BLR means it responds
very rapidly to changes in the ionizing continuum, whereas the
low-density, extended scale (10s to 100s of parsecs) of the NLR leads to
a slower response by many orders of magnitude: the BLR sees the central
continuum as it was days ago, whereas the NLR would see it as it was
decades to centuries ago.  Historically, at least a few Seyfert 1s have been
observed during extremely low continuum states in which the broad
components of the Balmer lines have practically vanished for a brief
time, e.g., NGC\,4151 \citep{pp84}, NGC\,1566 \citep{a85}, and Mrk\,1018
\citep{c86}.

Another possibility is that there are natural limits to the existence of
a BLR.  \citet[also \citealt{n03}]{n00} has proposed a scenario in which
the broad lines arise in a disk wind that requires a minimum accretion
rate, and the BLR naturally vanishes in low-luminosity objects that fall
below the minimum accretion-rate threshold.  \citet{laor03} has likewise
proposed that the BLR may vanish in low-luminosity AGNs, based on
extrapolations of empirical scaling relations for luminosity, BLR size
scale, and line widths that have been discovered in studies of AGN
spectral variability \citep[see][]{peterson04}.

Whatever the root cause, at first sight a true Seyfert 2 would be
virtually indistinguishable from a Seyfert 2 with a hidden BLR and thus
hard to identify without a fair amount of supplementary data.  If the
picture that they are related to low accretion-rate objects is correct,
they would be expected to be relatively rare and primarily identifiable
only among nearby AGN.  Thus far, really convincing true Seyfert 2s have
proven elusive. This paper presents an attempt to confirm true Seyfert 2
candidates from the local, spectroscopically-identified Seyfert 2
population by using X-ray measurements to determine if an
obscuring medium is present.

This paper is organized as follows: \S\ref{sec:sampsel}
describes how the \textit{Chandra} targets were selected;
\S\ref{sec:datan} describes the observations and analysis of the data,
with individual targets discussed in \S\ref{sec:ic3}--\ref{sec:n54}; the
results are discussed in \S\ref{sec:discus}; a summary is presented in
\S\ref{sec:sum}.

\section{Sample selection}
\label{sec:sampsel}

Selection of candidate true Seyfert 2 nuclei proceeds by a process of
elimination.  We started with a sample of local
($v<3000$\,km\,s$^{-1}$) AGN taken from galaxies in the CfA
Redshift sample \citep{hb92,om93} and the Revised Shapley-Ames (RSA)
Catalog \citep{st87} as described by \citet{rm99}.  Previous ``structure
mapping'' image processing of a subset of these galaxies \citep{pm02}
showed that a number of the Seyfert 2s have bright, unresolved star-like
nuclei with no apparent evidence of host-galaxy scale absorption
\citep[e.g., as seen in some Seyfert 2s by][]{mgt98}.
We start with these to eliminate those nuclei that are obscured by
kiloparsec-scale dust lanes in the host galaxies.  Of the remaining
``stellar'' nuclei Seyfert 2s, we eliminate all of those with previous
spectropolarimetric detections of broad H$\beta$ or H$\alpha$ emission
lines, and thus archetypal hidden Seyfert 1s.

The final selection employed long-slit optical spectrophotometry of the
nuclei acquired with the \textit{STIS} instrument on \textit{HST} as
part of a separate, on-going study of Seyfert 2s with unobscured,
stellar nuclei \citep{pip06}.  Of the nine Seyfert 2s without broad H$\alpha$
emission lines seen in direct or polarized light, four have red continua
and narrow Balmer emission-line ratios consistent with significant
line-of-sight extinction, and thus were rejected from further
consideration.  This left us with a final sample of five Seyfert 2s ---
IC\,3639, NGC\,3982, NGC\,5283, NGC\,5427, NGC\,5347 --- all of which
have blue featureless continua; rich, high-excitation spectra with
highly-ionized species like [\ion{Fe}{10}], [\ion{Fe}{7}] and
[\ion{Ne}{5}]; and most of the emission arising from an unresolved
region within $\sim20$ pc of the active nucleus.

Our procedure was therefore to identify true Seyfert 2 candidates by
eliminating Seyfert 2s with substantial host galaxy obscuration (using
morphological data), hidden BLRs (using spectropolarimetry), and
unresolved obscuration on small scales (using the presence of blue
nuclear continua and strong, high-excitation lines).
There were no archival X-ray data available to use at the time that
would permit an assessment of the X-ray absorption arising on much
smaller scales, so we applied for and were granted time during
\textit{Chandra} AO5 to acquire observations of 4 of these: IC\,3639,
NGC\,3982, NGC\,5283, NGC\,5427.  Observations of NGC\,5347 were awarded
to another program \citep{lhkwz06}.

\section{Observations and Data Analysis}
\label{sec:datan}

The four targets were observed between November 2003 and March 2004 for 10 ks
each, resulting in about 9 ks of usable exposure time on each object.
Details of the observations are given in Table~\ref{tab:obs}. Each
observation used the S3 chip in ACIS in $1/8$ sub-array mode. All of the
targets were detected.

The data were processed using version 3.1 of the Chandra Interactive
Analysis of Observations (CIAO) package. Level 2 event lists were
re-processed with observation-specific bad pixel files. The event lists
were then converted to FITS images and the CIAO wavelet source detection
tool \textit{wavdetect} was run to determine source positions. For each
observation, source counts were extracted from a circular aperture
centered on the \textit{wavdetect} source position and with a radius of
4.67 pixels (or 2.3\arcsec, $1.3$ times the on-axis 95\%
encircled-energy radius at 1.5 keV on ACIS-S). This aperture ensured
that virtually all photons from a point source would be
included. Background counts were taken from an annulus with an inner
radius of twice, and an outer radius of five times, the source circle
radius, after excising any point sources that fell within the
annulus. While it is true that diffuse extended emission from the host
galaxy, if any, would have been included in the background, it does not
affect our analysis since we are concerned with the nucleus and the much
brighter circumnuclear emission. Counts were extracted in the 0.3 -- 8.0
keV (Broad), 2.5 -- 8.0 keV (Hard) and 0.3 -- 2.5 keV (Soft) bands. A
hardness ratio (HR) was defined as $HR = (H-S)/(H+S)$, where $H$ and $S$
are the counts in the hard and soft bands, respectively. Source counts
and hardness ratios are given in Table~\ref{tab:det}. Further processing
on individual objects is described in the sections that follow. In
general, for the two objects where the data are of sufficient quality
for spectral fitting, unbinned pulse invariant (PI) spectra in the
energy range 0.3 -- 8.0 keV were created as described by the applicable
CIAO ``Science Thread'' at the
website\footnote{\url{http://cxc.harvard.edu/ciao/threads/pieces/}} of
the Chandra X-Ray Center (CXC).  Spectral fitting was performed using
the CIAO tool \textit{Sherpa}. Spectral fits are summarized in
Table~\ref{tab:spec}.

Though the soft emission in the spectra we fit is likely dominated by
emission lines, our spectra have too low signal-to-noise ratio to
identify lines or blends of lines. The best that can be done is to fit
the upper envelope of the lines, and even this is not very well
constrained. Therefore, for the sake of simplicity and ease of fitting
we use a thermal bremsstrahlung model for the soft component, rather
than a more physically plausible but multi-parameter model like MEKAL.

\subsection{IC\,3639}
\label{sec:ic3}

\begin{figure}[p]
\plottwo{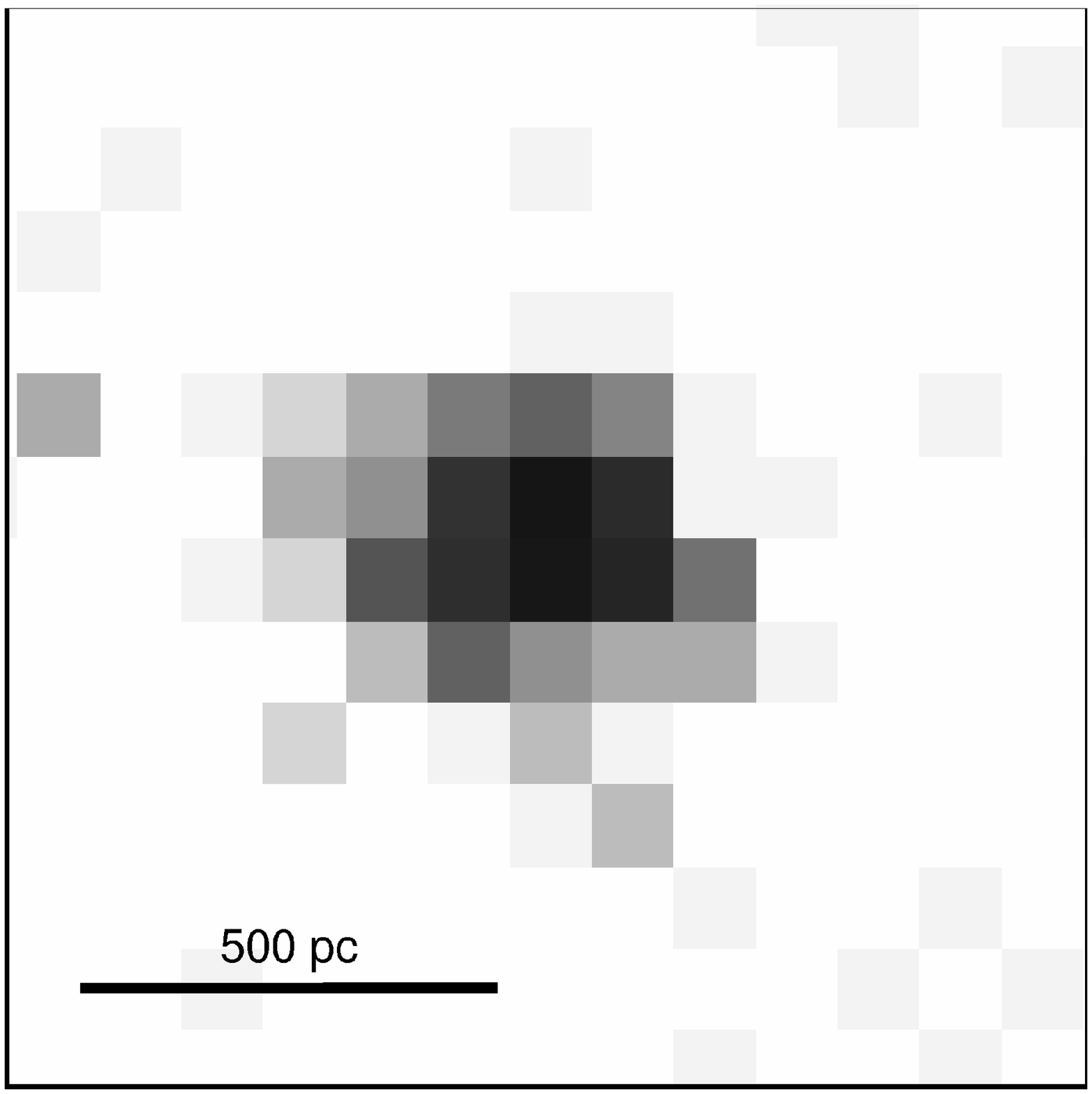}{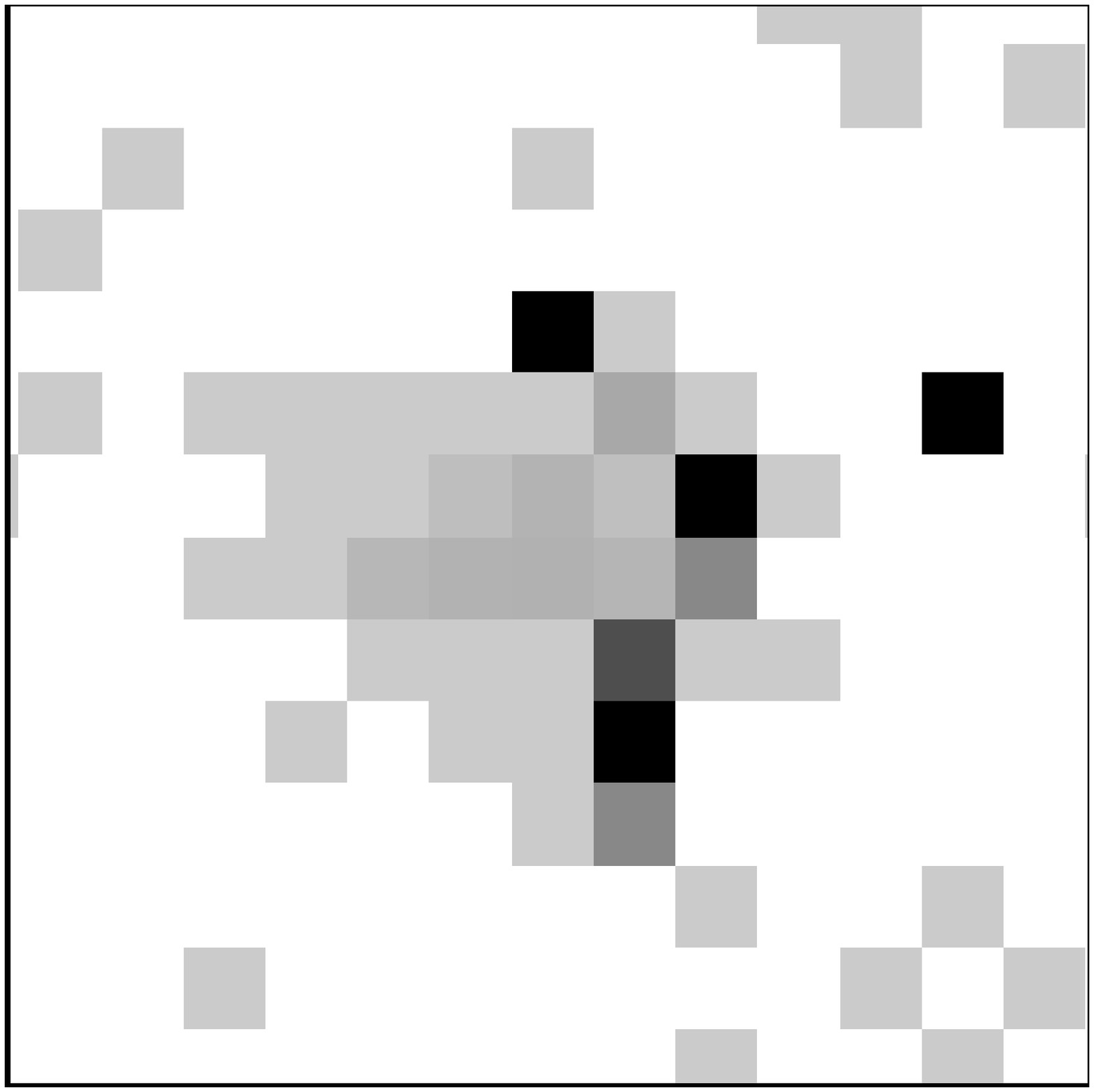}
\caption{\textit{Left: Chandra\/} image of the nucleus of IC 3639 (283
counts detected). There is extended circumnuclear emission in addition
to the point source. The solid black bar in the lower left represents a
projected distance of 500 pc. \textit{Right:\/} Hardness ratio (HR;
defined in \S\ref{sec:datan}) map of the nucleus, showing that most of
the emission is in the soft band. Each pixel is colored according to the
hardness ratio of the photons in that pixel. The lightest pixels show HR
$= -1$, with pixels becoming darker as the HR increases; the darkest
pixels have HR $= +1$. Both panels are 6.5\arcsec on a side. North is up
and East is to the left in both panels.}
\label{fig:ic3post}
\end{figure}

The nucleus of IC\,3639 is clearly detected, with 283 counts. As can be
seen from Fig.~\ref{fig:ic3post}, there is an extended
component to the emission and this extended emission is soft.

\begin{figure}[p]
\plottwo{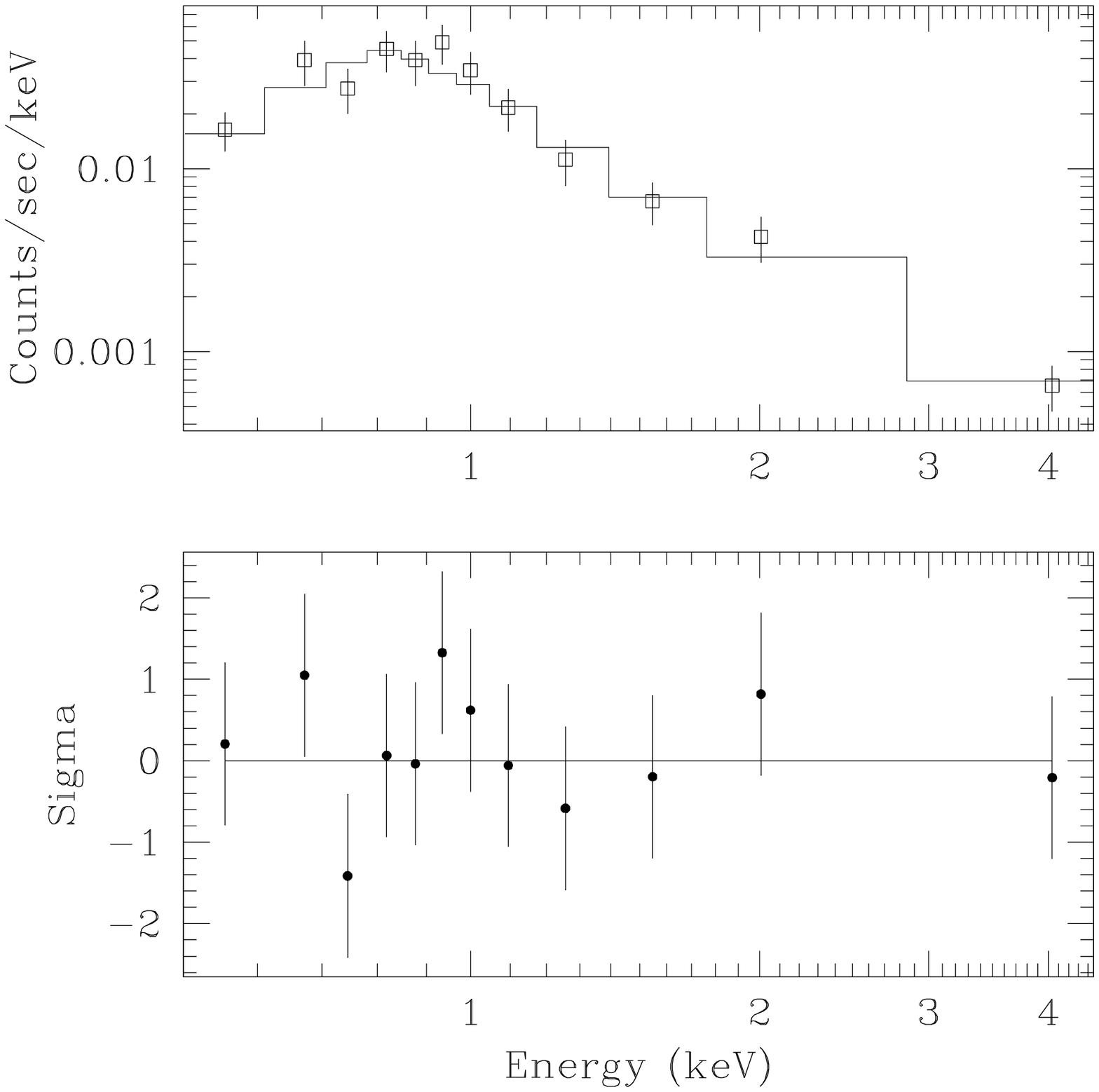}{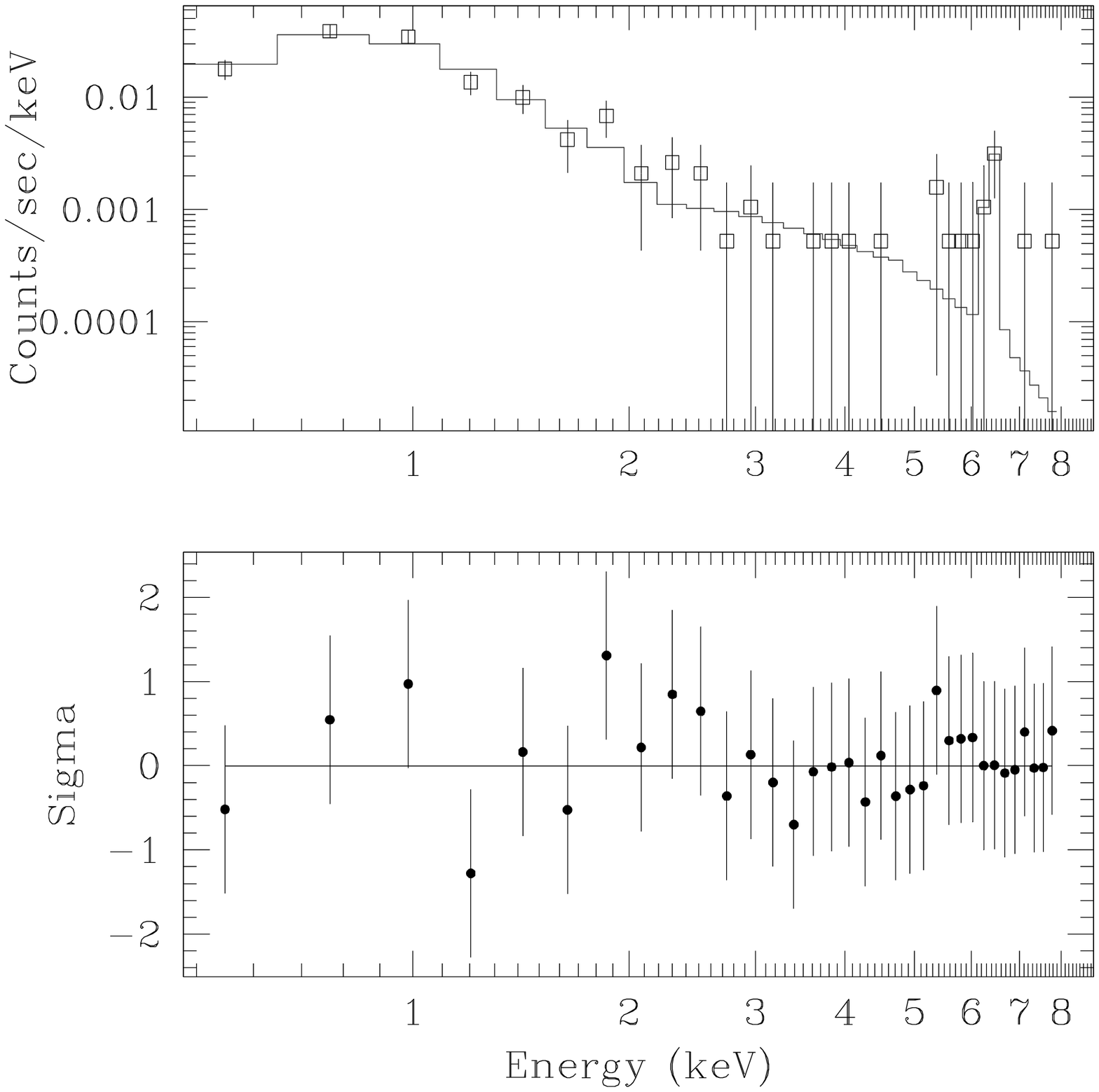}
\caption{Spectrum of IC 3639. The data include both the point source and
the extended component. \textit{Left:\/} Spectrum fit to an
absorbed power law plus thermal bremsstrahlung model. In the upper
panel, the squares with the error bars are the data and the solid line
the best-fit model, which is a power-law with photon index $\Gamma =
2.61 \pm 0.60$ plus thermal bremsstrahlung with a characteristic
temperature given by $kT = 0.11 \pm 0.03$ keV, both absorbed by an
equivalent hydrogen column density $N_{\mathrm H} = (8.9
\pm 2.4)\times 10^{21}$ cm$^{-2}$. The lower panel shows the residuals
from the fit. \textit{Right:\/} Spectrum fit to an absorbed bremsstrahlung plus
reflected power law model, with an emission line at $6.38 \pm 0.06$
keV. The best-fit parameters are as follows: $N_{\mathrm H} = (4.2 \pm
0.4)\times 10^{21}$ cm$^{-2}$, $kT = 0.20\pm 0.02$ keV, $\Gamma =
2.14\pm 0.79$. The nominal line equivalent width ($\sim 500$ eV) and the
fraction of reflected light in the spectrum ($R \sim 0.6$) are
reasonable but totally unconstrained by the fit. The lower panel shows
the residuals from the fit.} 
\label{fig:ic3allspec}
\end{figure}

The spectrum was extracted using the circular aperture described
above. Our extracted nuclear spectrum includes the extended component
because there are not enough counts to fit separate spectra to the
nuclear and extended components. Before fitting, the source spectrum PI
channels were binned so that each bin had at least 20 counts. The low
number of counts did not justify the extra parameters that would be
introduced by using models that have different absorbers for the
nuclear and extended emission, and therefore we assumed a single
absorber for both components. Since the extended soft emission is
included, it is not surprising that a simple, absorbed power-law model
(\texttt{powlaw1d} and \texttt{xswabs} in
\textit{Sherpa}) does not fit the spectrum well ($\chi_\nu^2 =
3$)\footnote{$\chi_\nu^2$ is the $\chi^2$ per degree of
freedom.}. Adding a thermal bremsstrahlung component (\texttt{xsbremss})
with $kT = 0.11 \pm 0.03$ keV provides an acceptable fit ($\chi^2_\nu =
0.9$). The resulting best-fit values for the power-law
index\footnote{This is a photon index. The number of photons as a
function of energy is assumed to be given by $N(E) \propto
E^{-\Gamma}$.} and column density are $\Gamma = 2.61 \pm 0.60$ and
$N_{\mathrm H} = (8.9\pm 2.4)\times 10^{21}$ cm$^{-2}$,
respectively. The spectrum along with the best-fit model and residuals
are shown in Fig.~\ref{fig:ic3allspec}. The spectrum suggests that we
are probably detecting an AGN with intrinsic power-law emission and
moderate to heavy absorption. The absorbing column density is not well
constrained due to the presence of the extended, soft component.

If we use bins of constant width, rather than bins with a minimum number
of counts, then there is evidence for an emission line at $\sim 6.4$
keV, albeit at low signal-to-noise level. We therefore tried a fit with
a reflection component. A thermal component with $kT = 0.20\pm 0.02$ keV
is still required. The best-fit position of the line is at $6.38\pm
0.07$ keV. The equivalent width of the line has a best-fit value of 500
eV but is unconstrained by the fit. Similarly, the reflection parameter,
which describes how much of the observed emission can be attributed to
reflection, is unconstrained but has a best-fit value of $0.6$. The
right panels of Fig.~\ref{fig:ic3allspec} show the fit, consisting of
thermal bremsstrahlung and a reflected power law, both absorbed, plus an
unresolved line.

The model flux in the bremsstrahlung component (0.5--2 keV, but
effectively 0.5 -- 10 keV, since there is almost no flux between 2 and
10 keV) is $7.1\times 10^{-12}$\,\ecs\ after correcting for absorption
($6.3 \times 10^{-14}$\,\ecs\ observed), which implies a luminosity
$L_{\mathrm{0.5-2\,keV}} = 1.9\times 10^{42}$\,\es. The
absorption-corrected AGN luminosities according to the model are
$L_{\mathrm{2-10\,keV}} = 1.7\times 10^{40}$ erg s$^{-1}$ and
$L_{\mathrm{0.5-10\,keV}} = 5.3\times 10^{40}$\,\es, or about 3\% of the
luminosity of the extended component.

\subsection{NGC\,3982}
\label{sec:n39}

\begin{figure}[p]
\plottwo{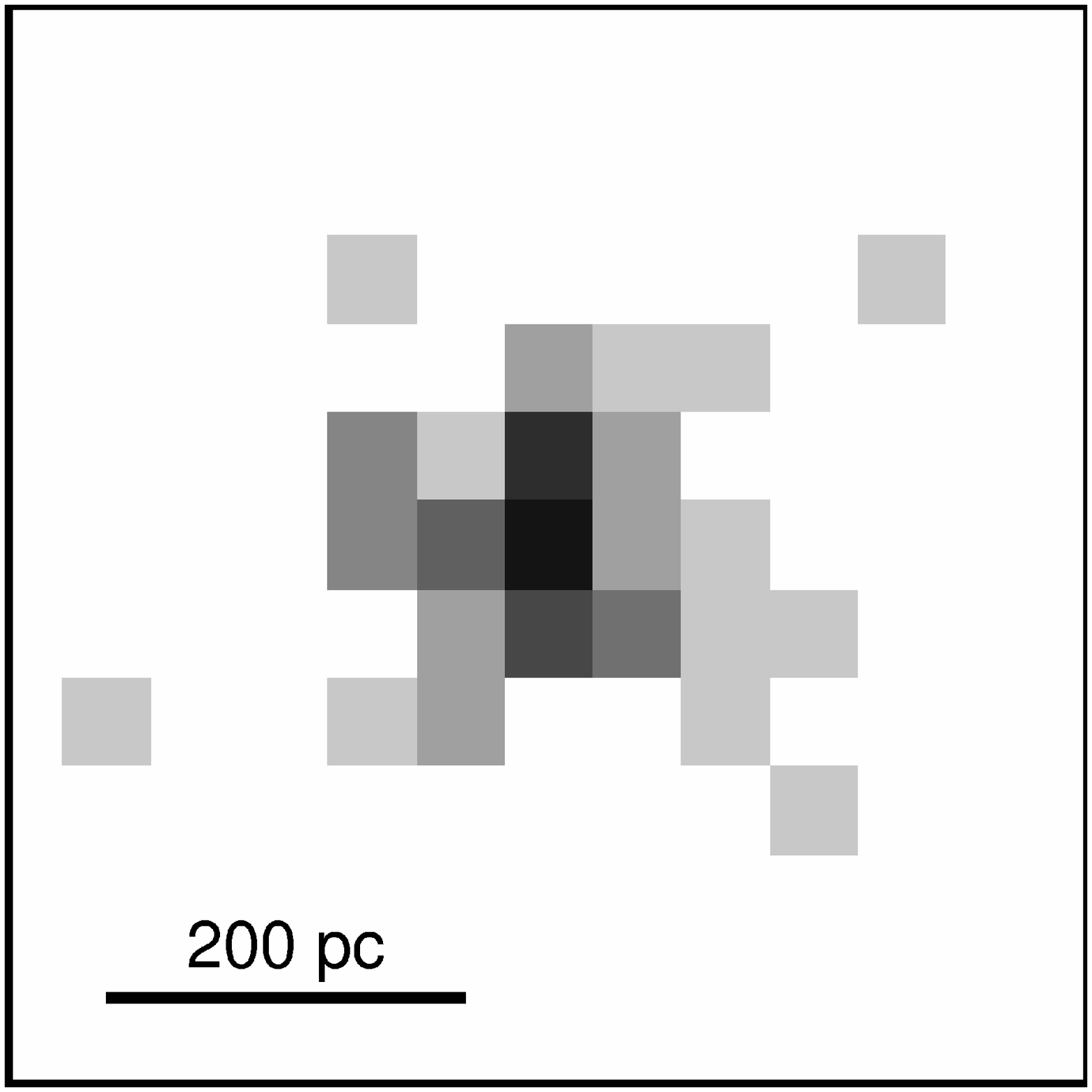}{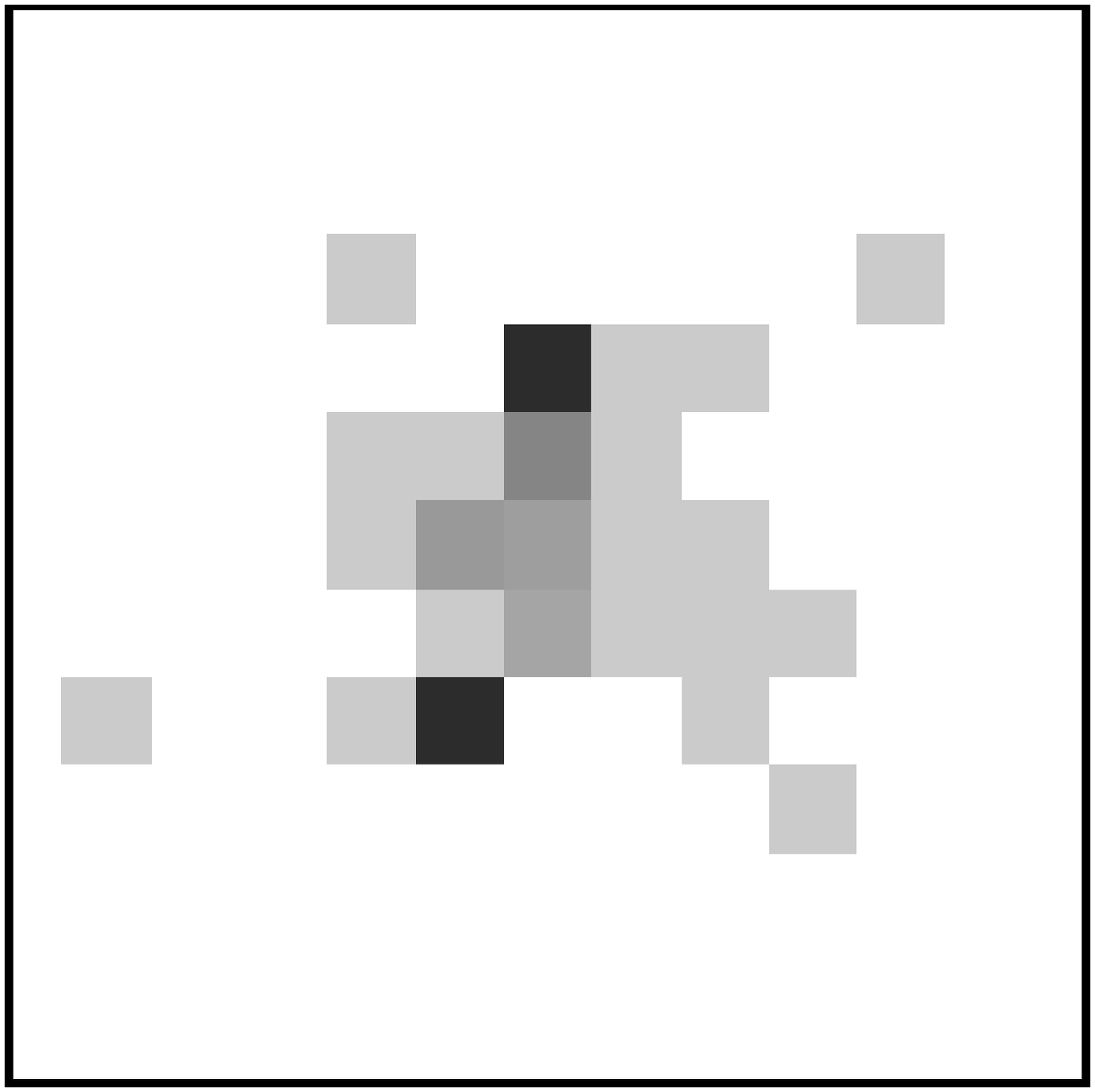}
\caption{\textit{Left: Chandra\/} image of the nucleus of NGC
3982 (73 counts detected). The solid black bar in the lower left represents a
projected distance of 200 pc. \textit{Right:\/} Hardness ratio (HR;
defined in \S\ref{sec:datan}) map
of the nucleus, showing that most of the emission is in the soft
band. Each pixel is colored according to the hardness ratio of the
photons in that pixel. The lightest pixels show HR $= -1$,
with pixels becoming darker as the HR increases; the darkest pixels have
HR $= 0$. Both panels are 6\arcsec on a side.  North is up
and East is to the left in both panels.}
\label{fig:n39post}
\end{figure}

\begin{figure}[p]
\plottwo{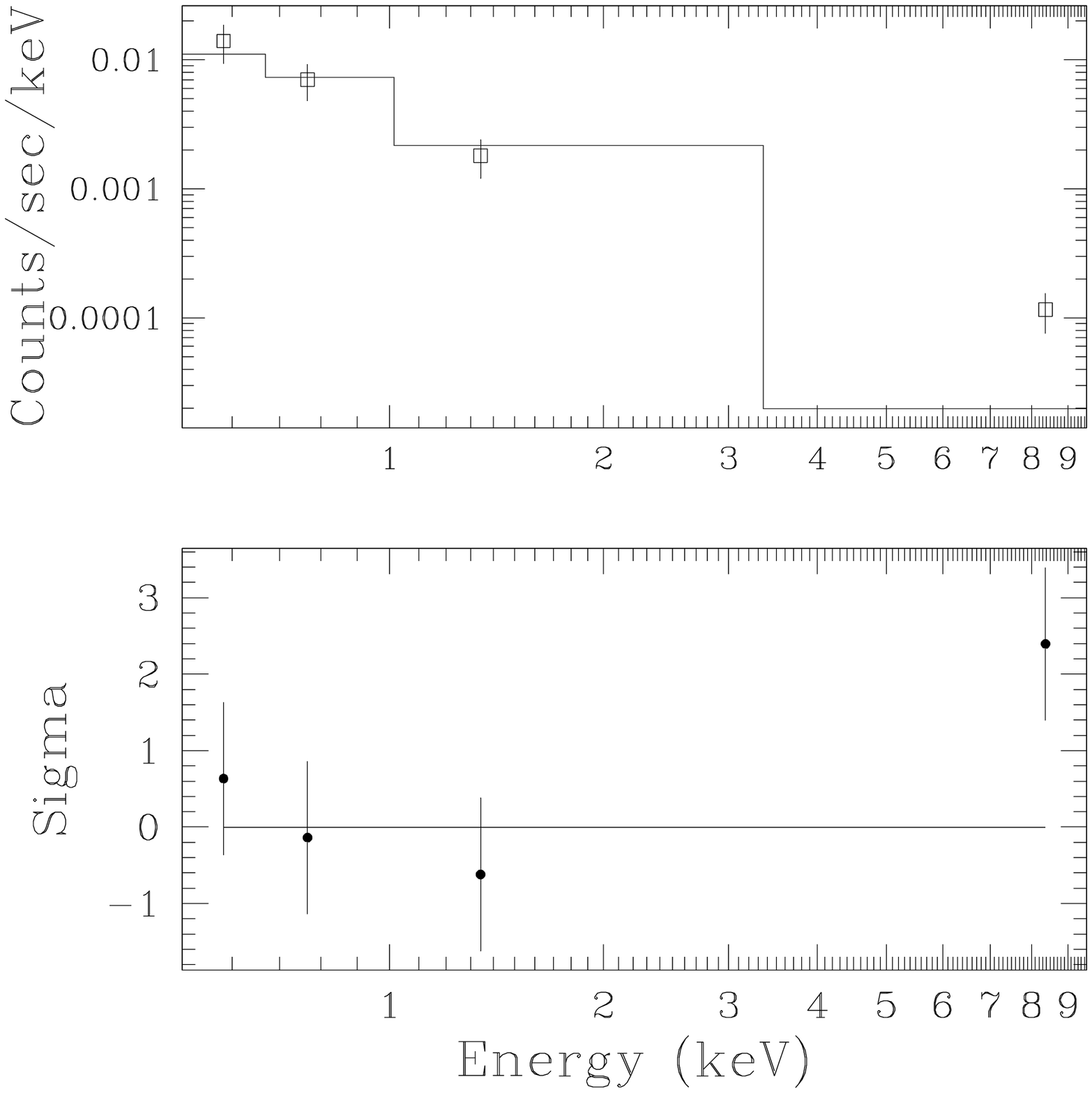}{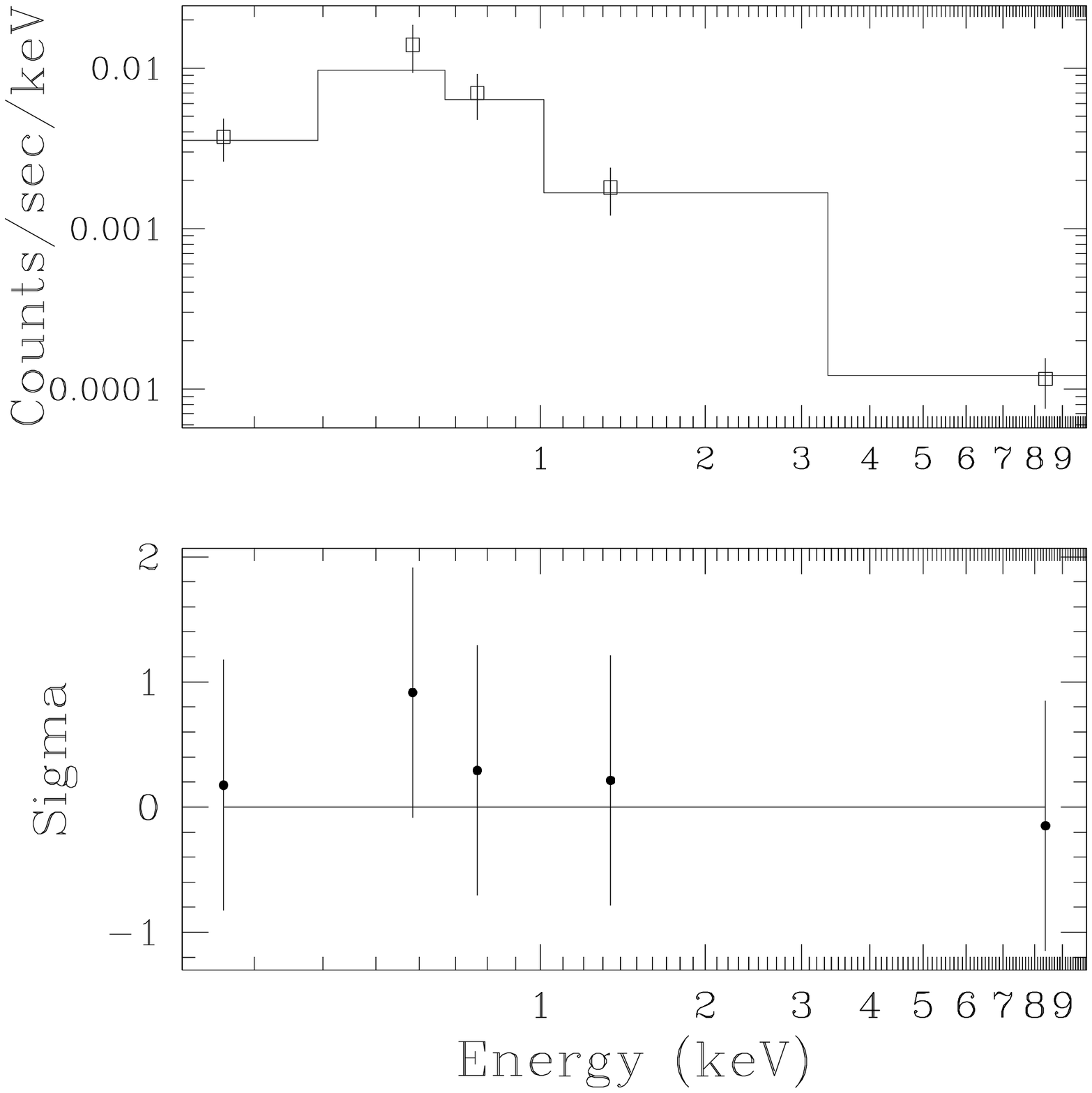}
\caption{Spectrum of NGC 3982 compared to two models. \textit{Left:\/} The squares show the binned data points and the solid line shows the ``best-fit'' power law,
$\Gamma = 3.7 \pm 0.9$, with only Galactic absorption ($1.23 \times
10^{20}$ cm$^{-2}$). The lowest-energy point, seen in the right panel, was not used in estimating the slope of the power law. The lower panel shows the
residuals. \textit{Right:\/} The squares show the binned data points and
the solid line shows an absorbed thermal bremsstrahlung plus
power law model, where absorption is a free parameter but $\Gamma$ and
$kT$ are kept fixed. The parameter values are $kT = 0.13$ keV and $\Gamma
= 1.0$ with $N_H= (2.34\pm 0.85)
\times 10^{21}$ cm$^{-2}$. The lower panel shows the residuals.}
\label{fig:n39spec}
\end{figure}

The nucleus of NGC\,3982 was detected with 73 counts.
Figure~\ref{fig:n39post} shows the \textit{Chandra} image and a hardness
ratio map. Most of the emission is in the soft band. The source does not
appear extended. The small number of counts precludes any detailed
spectral fitting, but we attempted several two-parameter fits (one interesting
parameter and its normalization) with the column density fixed at the
Galactic value towards this target ($1.23
\times 10^{20}$ cm$^{-2}$). After binning the PI channels so that there
were at least 15 counts in each bin, a power-law model gives a best-fit
value of $\Gamma = 3.7 \pm 0.9$. However, a single power law does not
provide a good fit ($\chi^2_\nu \sim 3.3$) as can be seen in the left
panels of Fig.~\ref{fig:n39spec}.

Considering the possibility that the entire emission seen could be from
circumnuclear gas, such as is present in IC\,3639, the spectrum was also
fit with a thermal bremsstrahlung model and a model with both power-law
and thermal components.  The former gives a best-fit value of $kT =
0.54\pm 0.15$ keV, with $\chi^2_\nu = 3.3$. The latter fit was performed
by first fitting a bremsstrahlung model to the data points below 2 keV,
then freezing the model and fitting a combined thermal plus power law
model to data in the full energy range, 0.3--8.0 keV. This produces a
nominally better fit, with $\chi^2_\nu = 1.6$, but fails to constrain
any of the fit parameters. The best-fit values are $kT = 0.48$ keV and
$\Gamma = -1.6$.  The above fit was also re-done with absorption as a
free parameter, but keeping the power law index and temperature of the
thermal emission fixed. For $\Gamma = 2$, the fit returns $N_H = (2.2\pm
0.9)\times 10^{21}\, \mathrm{cm}^{-2}$ and $kT = 0.13$ keV.  The
best-fit values were obtained by varying the (fixed) values of $\Gamma$
and kT and re-fitting, giving $N_H = (2.3\pm 0.9)\times 10^{21}$
cm$^{-2}$, $\Gamma = 1.0$ and $kT = 0.13$ keV. This fit is shown in the
right panels of Fig.~\ref{fig:n39spec}.

\subsection{NGC\,5283}
\label{sec:n52}

\begin{figure}[p]
\plottwo{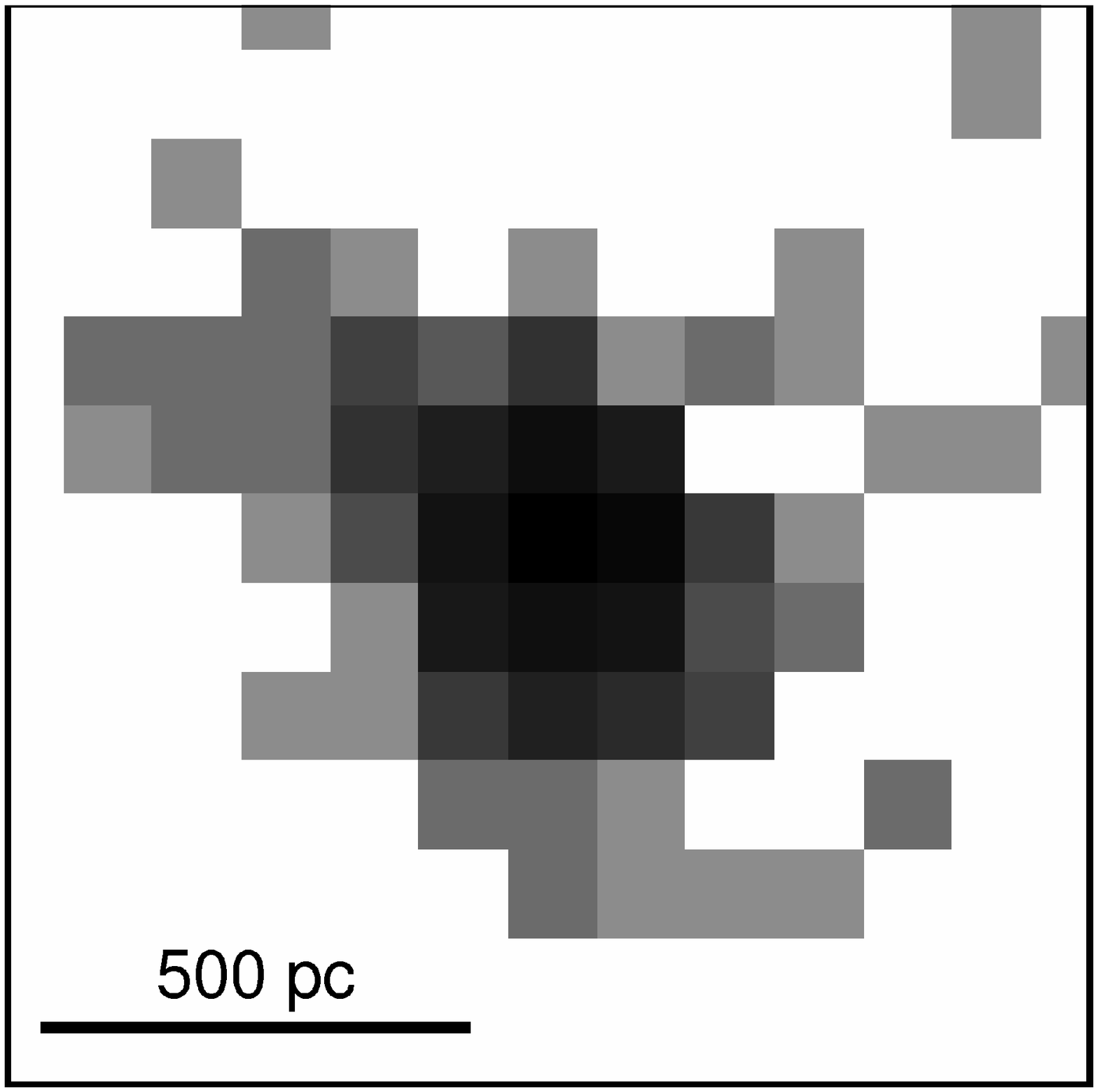}{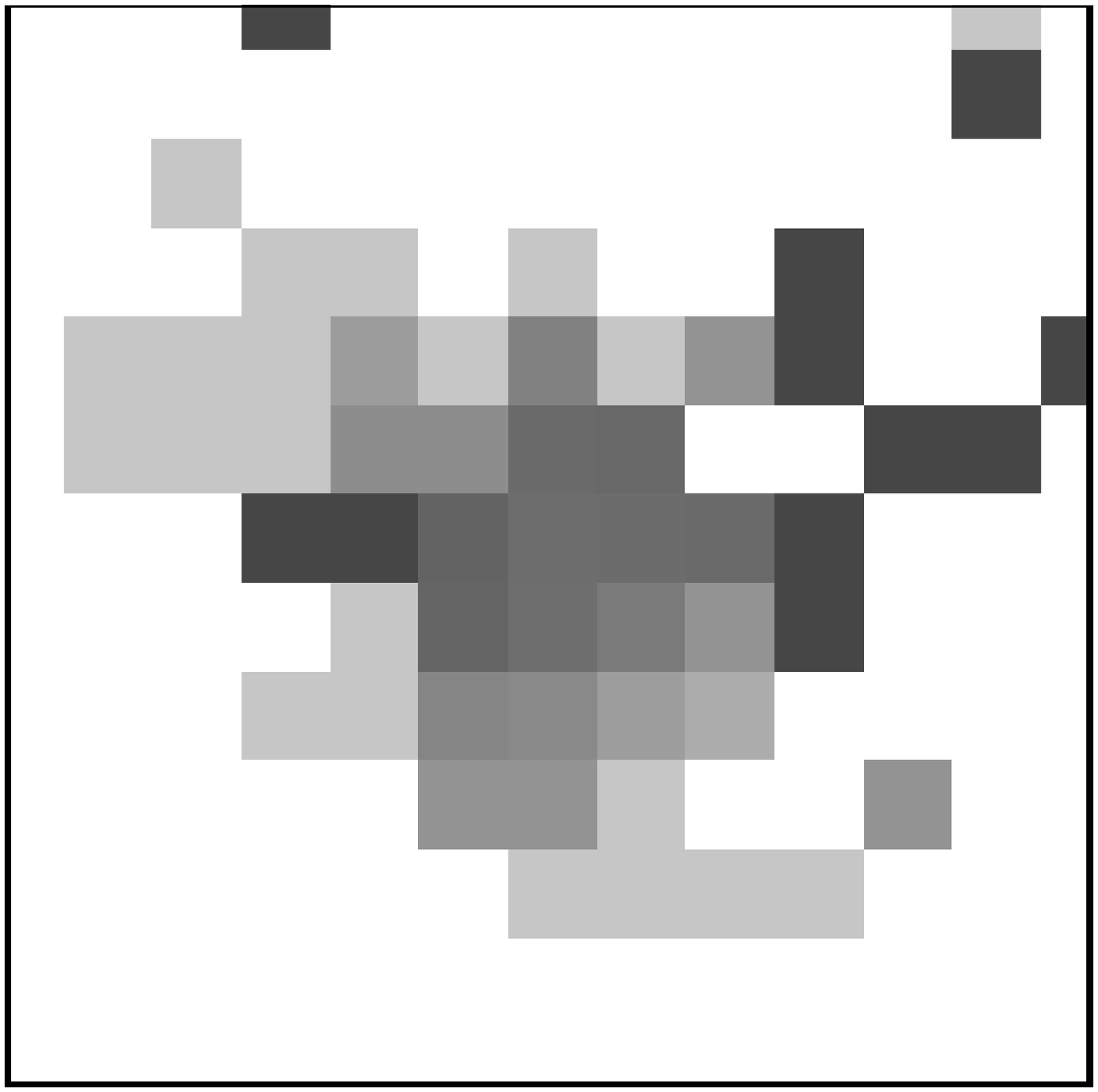}
\caption{\textit{Left: Chandra\/} image of the nucleus of NGC
5283 (454 counts detected). The solid black bar in the lower left represents a
projected distance of 500 pc. \textit{Right:\/} Hardness ratio (HR;
defined in \S\ref{sec:datan}) map
of the nucleus, showing that most of the emission is in the hard
band. However, the most extended emission is also the softest. Each
pixel is colored according to the hardness ratio of the photons in that
pixel. The lightest pixels show HR $= -1$, with pixels
becoming darker as the HR increases; the darkest pixels have HR
$=+1$. Both panels are 6\arcsec on a side. North is up
and East is to the left in both panels.}
\label{fig:n52post}
\end{figure}

\begin{figure}[p]
\plotone{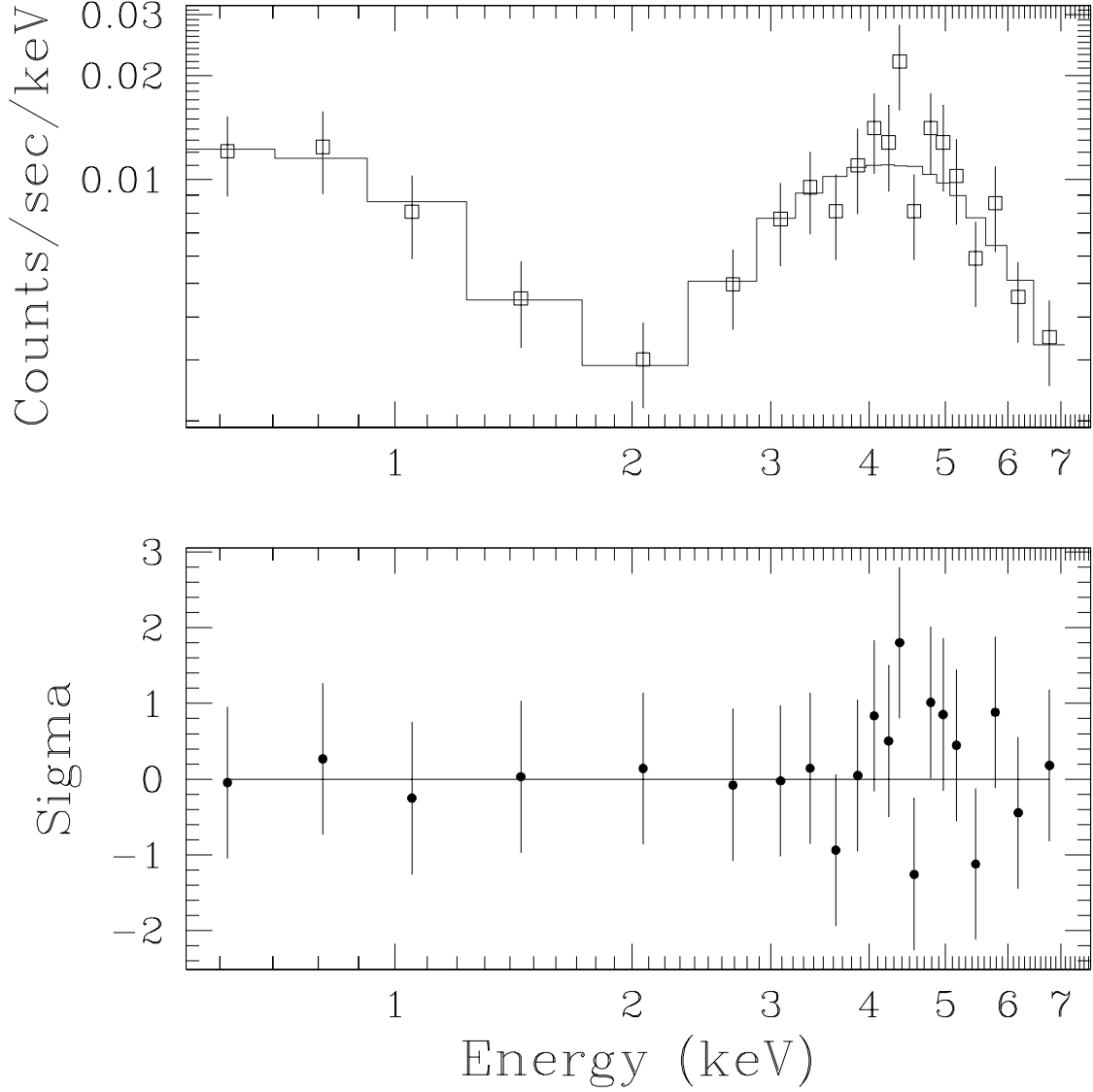}
\caption{Spectrum of NGC 5283 including both the point source and extended
emission. The squares show the binned data points and the solid line the
best-fit model. The model consists of an absorbed power law ($\Gamma =
0.76 \pm 0.65$, $N_{\mathrm H} = [7.3 \pm 3.0]\times 10^{22}$ cm$^{-2}$)
and thermal emission ($kT = 0.71\pm 0.27$ keV). The Galactic neutral
hydrogen column density is $N_{\mathrm H} = 1.84\times 10^{20}$
cm$^{-2}$. The lower panel shows the residuals from the fit.}
\label{fig:n52spec}
\end{figure}

NGC\,5283 is the brightest of the four targets, with 454 counts. The
\textit{Chandra} image (Fig.~\ref{fig:n52post}) clearly shows extended
emission to the northwest and southeast of the central source. This
object is different from the other three in that its hardness ratio is
positive, i.e.\ there is more emission in the hard band. It is similar
to IC\,3639 in that there is a central source surrounded by extended soft
emission --- in this case the point source is hard, and bright enough to
dominate the total emission. Spectral fits were performed with PI
channels binned so that each bin had at least 20 counts. The spectrum
(Fig.~\ref{fig:n52spec}) shows a power law at high energies (best-fit
$\Gamma = 1.46 \pm 0.33$), but with a turn-over at $\sim 4$ keV,
indicating strong absorption (best-fit $N_{\mathrm H} = [7.5 \pm 3.0]
\times 10^{22}$ cm$^{-2}$). There is a significant ``soft excess'' which
can be fit by a thermal bremsstrahlung model with best-fit $kT = 0.79
\pm 0.31$ keV. 

\begin{figure}[p]
\plotone{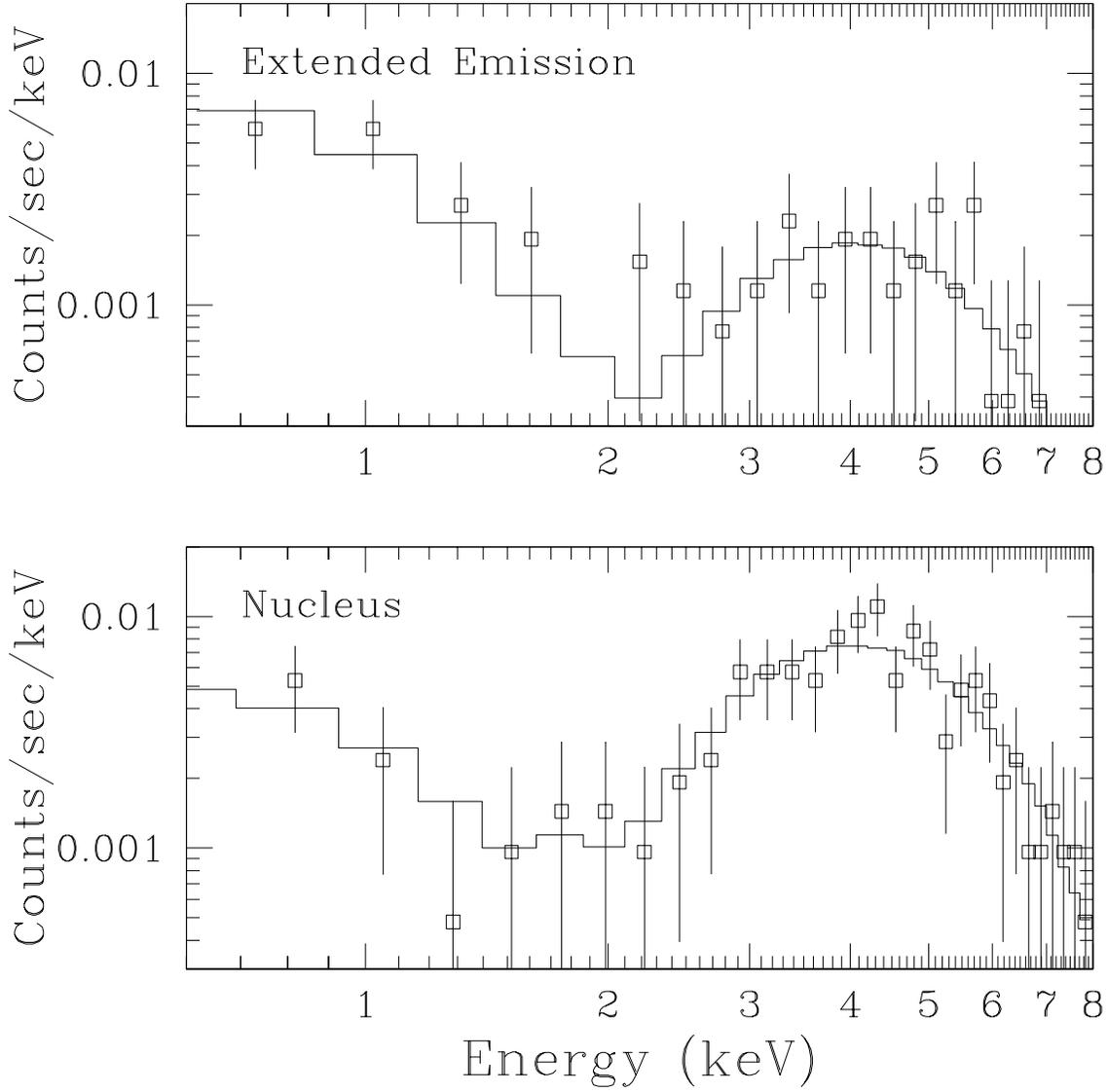}
\caption{Separate spectra of the core and surrounding region of the
nucleus of NGC 5283.  The top panel shows the spectrum of just the extended
emission, with the central $3\times 3$ island masked out. The bottom panel
shows the spectrum of just the central pixel. The spectra suggest that
the soft, thermal component is emission from the extended emission seen
in Fig.~\ref{fig:n52post}, while the central source provides the hard,
absorbed power law component.}
\label{fig:n52specsep}
\end{figure}

\begin{figure}[p]
\plotone{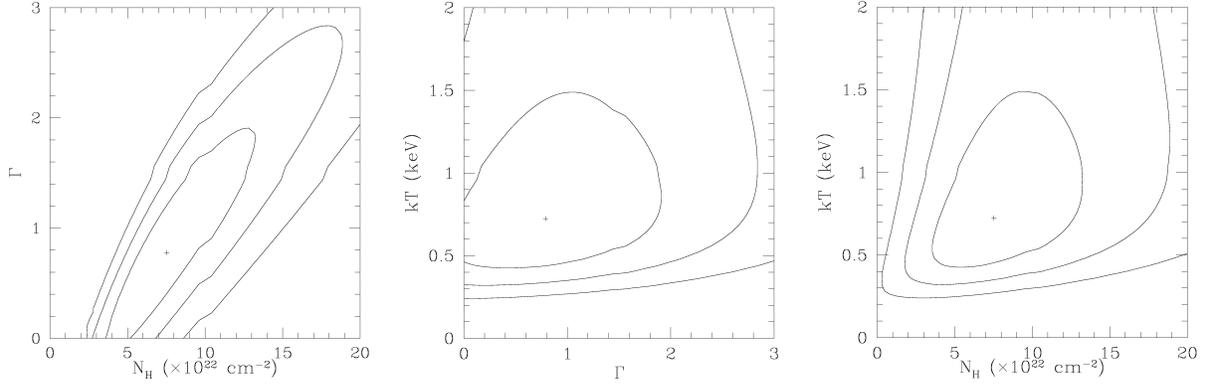}
\caption{Confidence region projections for parameter pairs for the NGC
5283 spectral fit of the point source and extended emission together. A
plus sign marks the best-fit value. Contours are drawn at the
$1,2,3-\sigma$ levels.}
\label{fig:n52conf}
\end{figure}

Most of the soft emission is from the extended feature seen in
Fig.~\ref{fig:n52post} and we therefore extracted spectra of this
emission and the nuclear emission separately. The top panel of
Fig.~\ref{fig:n52specsep} shows a spectrum of just the diffuse
emission. This was extracted from a 12-pixel $\times$ 8-pixel rectangle
which encompassed the entire extended emission, but with a $3\times 3$
island centered on the brightest pixel masked out. The bottom
panel shows a spectrum of just the central pixel. Of note about this
object is that \emph{no} fit was possible if the thermal component and
the power law were assumed to be absorbed by the same gas. The only
acceptable fits required a much lower column density for the material
absorbing the thermal emission than for the material absorbing the power
law emission.  Confidence regions for the spectral model parameters are
shown in Fig.~\ref{fig:n52conf}.

The best-fit model gives absorption-corrected nuclear luminosities of
$L_{\mathrm{2-10\,keV}} = 6.3\times 10^{41}$\,\es\ and
$L_{\mathrm{0.5-10\,keV}} = 7.1\times 10^{41}$\,\es, and the extended
emission has $L_{\mathrm{0.5-2\,keV}} = 9.9\times 10^{39}$\,\es. The AGN
is brighter by a factor of $\sim\! 65$.

\subsection{NGC\,5427}
\label{sec:n54}

\begin{figure}[p]
\epsscale{0.5}
\plotone{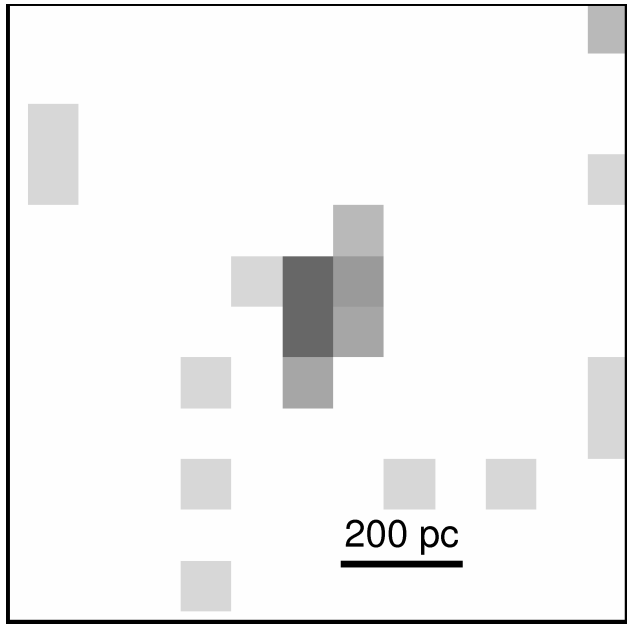}
\caption{\textit{Chandra\/} image of the nucleus of NGC
5427 (35 counts detected). The solid black line in the lower right
represents a projected distance of 200 pc. The image is 6\arcsec on a
side. North is up and East is to the left.}
\label{fig:n54post}
\end{figure}

\begin{figure}
\epsscale{1.0}
\plotone{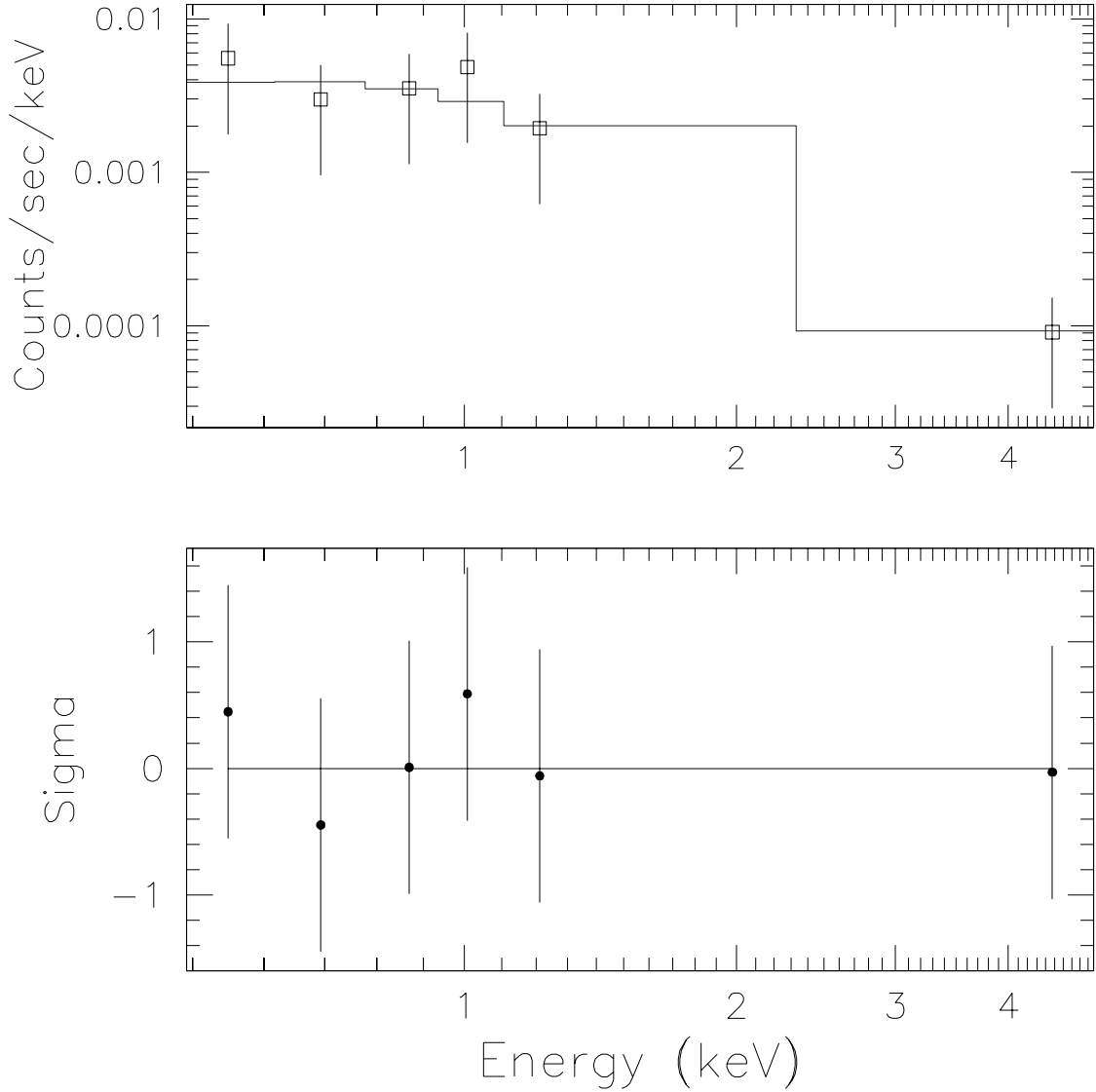}
\caption{Spectrum of NGC 5427, binned so that each bin has at least 5
counts. The solid line shows a thermal bremsstrahlung
model characterized by $kT \sim 0.7$ keV with Galactic absorption ($N_H
= 2.38\times 10^{20}$ cm$^{-2}$). The lower panel shows the residuals.}
\label{fig:n54spec}
\end{figure}

Figure~\ref{fig:n54post} shows the \textit{Chandra} image of the nucleus
NGC\,5427. Only 35 counts were detected, almost all (33) in the soft
band, giving it a hardness ratio of $-0.9$. If this source is similar to
the others, then the faintness and lack of hard emission suggest we are
detecting only the circumnuclear soft emission and the central point
source is heavily absorbed. Figure~\ref{fig:n54spec} shows the spectrum,
which has been binned so that there are at least 5 counts in each
bin. As with the other targets, a thermal bremsstrahlung model ($kT =
0.7$ keV) with Galactic absorption was ``fit'' to the data. However,
there are not sufficient data to discriminate among models. The data are
equally consistent with a model where the source is assumed to consist
of absorbed thermal bremsstrahlung and a power law.

\section{Discussion}
\label{sec:discus}

Although targeted as candidate true Seyfert 2s, or AGNs without BLRs, the
four objects studied here --- IC\,3639, NGC\,3982, NGC\,5283, and
NGC\,5427 --- all show evidence of obscuration when observed in X-rays.
They are therefore not good ``true'' Seyfert 2 candidates, though of
course the presence of obscuration does not imply the existence of a BLR
behind the obscuration. The picture that emerges is of moderate to high,
but not necessarily Compton-thick, obscuration of the nucleus, $N_H
\sim $ few $\times 10^{21}$ -- $10^{22}$ cm$^{-2}$, and extended
soft emission on hundreds-of-parsecs scales that may possibly originate
in the NLR.

The spectrum of IC\,3639 shows moderate absorbing column densities and
possibly a reflection component, with an Fe K$\alpha$ line detected with
low significance. The spectrum of NGC\,5283 clearly shows a power law
which is strongly absorbed below $\sim 4$ keV. The spectra of both
objects require a separate component to fit the soft emission. This
component is probably a result of the blending of multiple emission
lines from a photoionized gas \citep{bgc05}; however, the CCD spectra
analyzed here are not of high enough quality to constrain a
multi-parameter model like APEC or MEKAL, and the soft emission was equally
well fit by a thermal bremsstrahlung model and a pure
blackbody spectrum. In either case the characteristic temperature turns
out to be a few tenths of a keV. That the soft and hard components
originate in physically distinct regions is suggested by the hardness
ratio maps themselves (Figs.~\ref{fig:ic3post} and \ref{fig:n52post}),
which show harder emission surrounded by extended softer emission. In
the case of NGC\,5283, which has sufficient counts, this was shown more
definitively by fitting the spectra of the nucleus and the surrounding
emission separately. The harder, power law component clearly is stronger
closer to the nucleus, while the extended emission is dominated by the
soft component. The extended, soft component also has a lower obscuring
column than the nuclear emission.

The two fainter objects, NGC\,3982 and NGC\,5427, while too faint to make
spectral fitting possible, appear to be consistent with the above
picture. The observed emission is soft: NGC\,3982 has a hardness ratio of
$-0.7$ and NGC\,5427 does not even have a statistically significant
detection in the hard band. The faintness is not due to distance
--- these two objects are in fact nearer than the two brighter ones. All
four nuclei observed here have comparable [\ion{O}{3}] luminosities
\citep{w92a,nw95} and thus are expected to have comparable
$L_X$ \citep{hphk05}. These two fainter objects may therefore be
even more heavily obscured and we are primarily detecting soft X-ray
emission from hot, extra-nuclear gas. Is this gas itself
responsible for part of the absorption of the nucleus? While not clear
in IC\,3639, in NGC\,5283 the gas spectrum is consistent with only
Galactic absorption, whereas the nuclear spectrum requires an additional
column density of $N_H \sim 10^{22}$ cm$^{-2}$. In the \citet{laor03} model,
the innermost part of an NLR would become a warm absorber, with the
effect getting stronger for lower-luminosity AGN.

\begin{figure}
\plotone{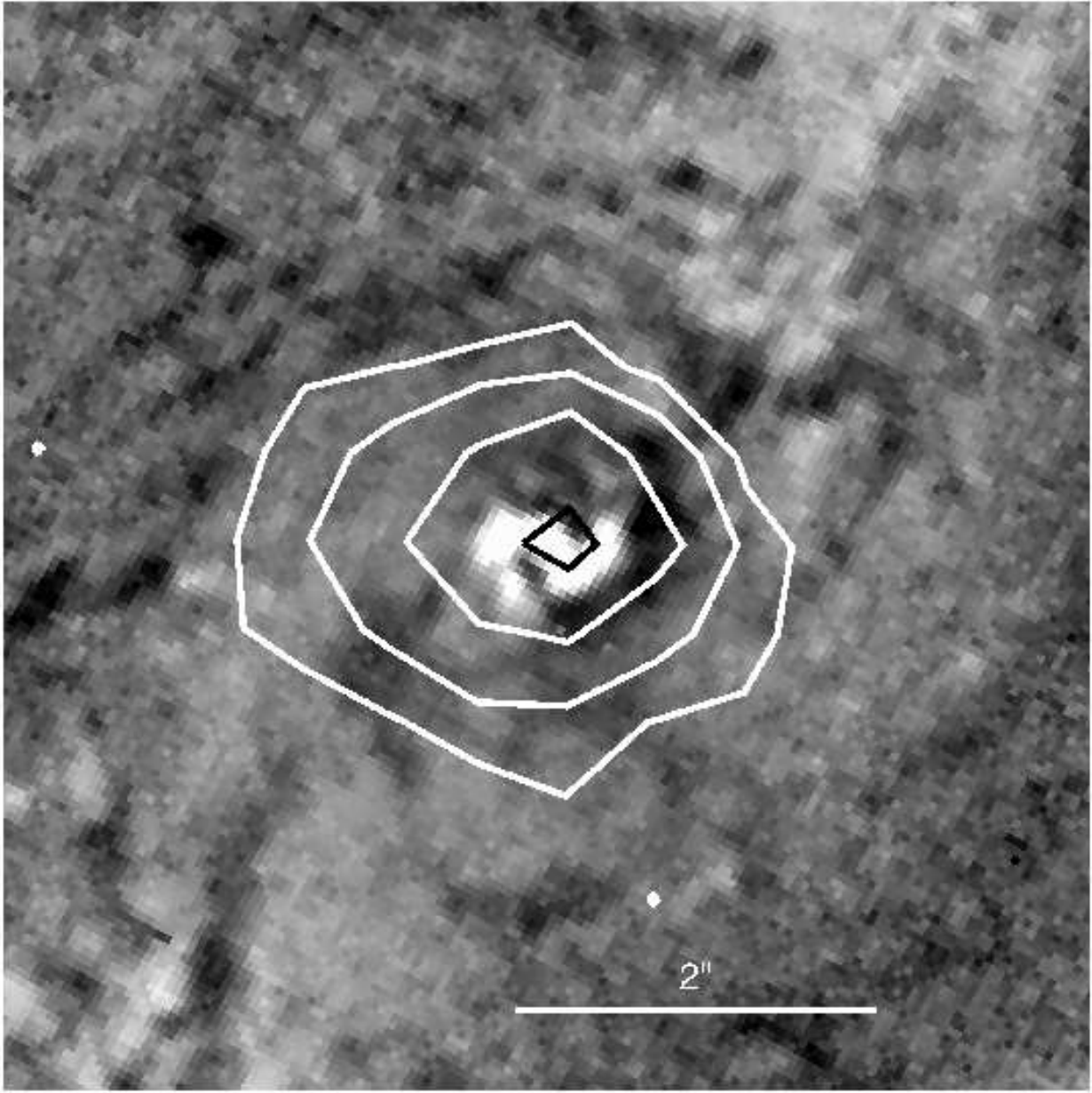}
\caption{Structure map of the nucleus of IC 3639 with X-ray contours
overlaid. The structure map is derived from an \textit{HST} image taken
using the F606W filter. The bright region near the center is the
narrow-line region. The X-ray contours show extension in the East-West
direction, as does the NLR. The image is 6\arcsec\ on a side, and the
white bar on the lower right represents 2\arcsec\ ($\sim\!400$ pc). The
contours are drawn at levels of 2.5, 6.3, 15.8 and 39.8 photons per
pixel. The innermost contour is plotted in black to enhance
visibility. North is up and East is to the left in the image.}
\label{fig:ic3hstxcont}
\end{figure}

\begin{figure}
\plotone{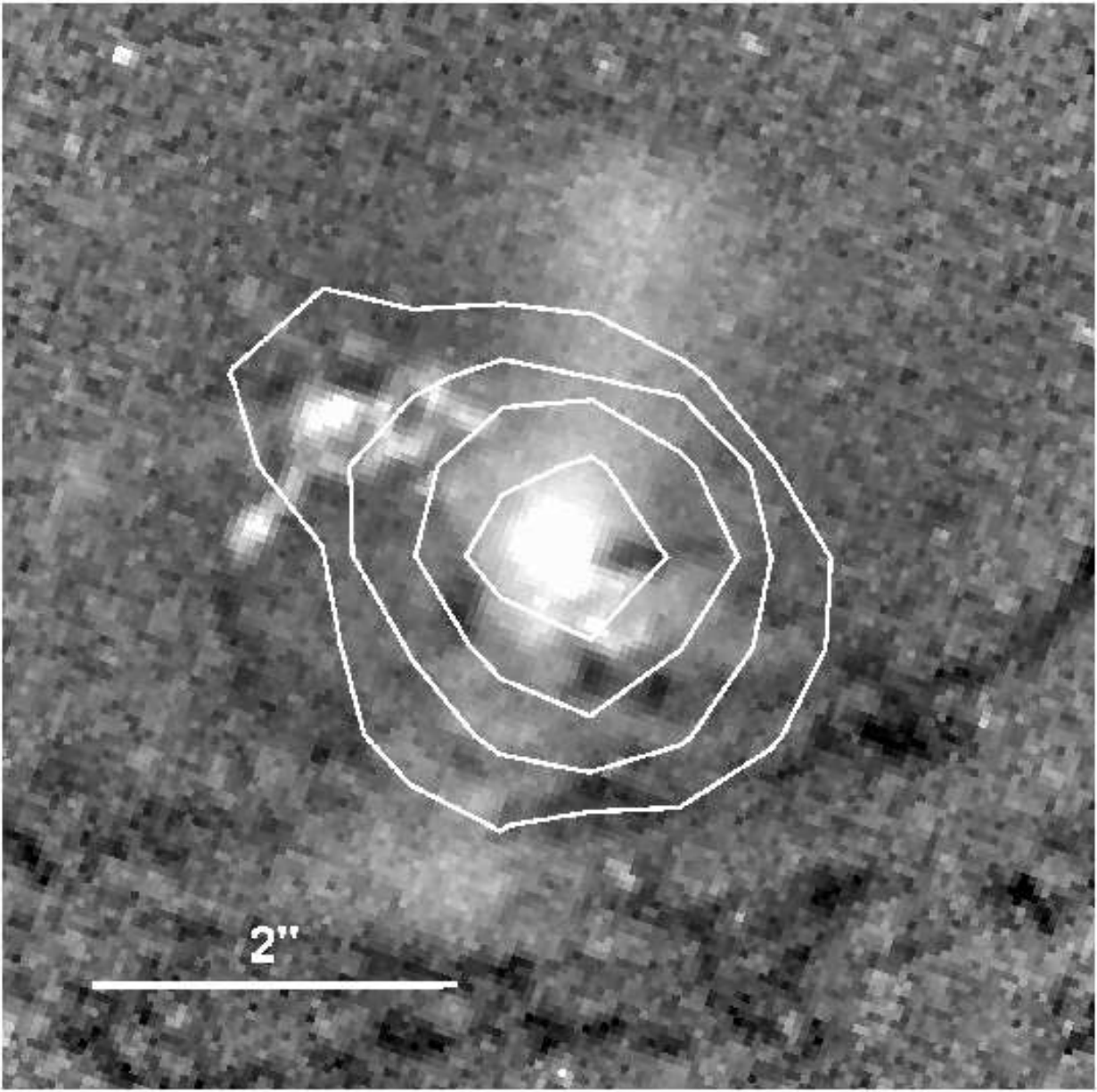}
\caption{Structure map of the nucleus of NGC 5283 with X-ray contours
overlaid. The structure map is derived from an \textit{HST} image taken
using the F606W filter. North is up and East is to the left in the
image. The sharp bright regions near the center show emission from the
narrow-line region. The X-ray contours show significant extension in the
same direction as the region of enhanced narrow-line emission seen on
the structure map. The diffuse bar-like structure running roughly
Northwest--Southeast is stellar. The image is 6\arcsec\ on a side, and
the white bar on the lower left represents 2\arcsec\ ($\sim\!450$
pc). The contours are drawn at levels of 2.5, 6.3, 15.8 and 39.8
photons per pixel.}
\label{fig:n52hstxcont}
\end{figure}

The gas emitting soft X-rays has the right physical scale
(few hundred parsecs) to be the gas that constitutes the NLR, and
\citet{bgc05} find that in their sample of Seyfert 2s the morphology of
the X-ray gas matches that of the NLR as determined by [\ion{O}{3}]
$\lambda 5007$ imaging. Figures~\ref{fig:ic3hstxcont} and
\ref{fig:n52hstxcont} show structure maps \citep{pm02} made from
wide-band (F606W) \textit{HST} images of IC 3639 and NGC 5283,
respectively, with X-ray contours overlaid. The NLRs are identifiable as
bright regions near the centers of the images, and the contours show
that the X-ray emission is extended in the same orientation as the
NLRs. NGC\,3982 and NGC\,5427, which do not show prominent, extended
X-ray emission, also do not show extended NLR emission in structure maps
\citep{pm02}. If the soft X-ray-emitting gas is really the NLR, then the
source of ionizing radiation is presumably the central engine, and not a
starburst. The soft X-ray (0.5--2 keV) luminosity of IC\,3639 inferred
from the spectral fit, $L_{0.5-2\mathrm{keV}} \sim 10^{42}$ erg
s$^{-1}$, would imply a star formation rate (SFR) of $\sim 100$ $M_\sun$
yr$^{-1}$ \citep{hhptc05}. This is an order of magnitude higher than the
known SFR of IC\,3639 derived from its infrared luminosity \cite[$\sim\! 
9\, M_\sun$ yr$^{-1}$,][]{dpkc02} and favors the central engine as the
ionizing source.  For NGC\,5283, $L_{0.5-2\mathrm{keV}} \sim 10^{40}$
erg s$^{-1}$, requiring an SFR of only a few $M_\sun$ yr$^{-1}$, which
is consistent with what is known based on its infrared luminosity
\citep{k98,pgre01}, and star formation cannot be ruled out, but the AGN
luminosity of $\sim 10^{42}$ \es\ could explain the ionized gas without
star formation.  For IC\,3639, the [\ion{O}{3}] to soft X-ray flux ratio
$F_{\lambda 5007}/F_{0.5-2\mathrm{keV}} \approx 5-6$, while the ratio
for NGC\,5283 is $F_{\lambda 5007} / F_{0.5-2\mathrm{keV}}
\approx 0.6$, a range similar to that seen in the \cite{bgc05}
sample. Higher quality spectra of similar, absorbed Seyferts indicate
that the soft X-ray emission consists of blended emission lines from a
photoionized gas
\citep[e.g.][]{bgc05,lhkwz06}. Thus, morphology, energetics, and spectra all
suggest that the NLR is the source of the soft X-ray emission seen in
at least some Seyfert 2s.

Though obscuration prevents any deductions about the presence or absence
of BLRs in these objects, we apply the \citet{laor03} model to our
objects to determine if the unobscured luminosities are still consistent
with his model. This model predicts a minimum bolometric luminosity (as
a function of black hole mass) needed to sustain a BLR. We obtained
published stellar velocity dispersions from \citet{w92a}, \citet{nw95}
and \citet{grea05}, [\ion{O}{3}] fluxes from \citet{w92a} and
\citet{nw95}, and used the conversion from \citet{hea04} to change
[\ion{O}{3}] luminosities to bolometric luminosities. A comparison with
Fig.~1 of \citet{laor03} shows that all four of our nuclei have
luminosities ($\log (L_\mathrm{bol}/\mathrm{erg}\,\mathrm{s}^{-1}) \sim
44,43,44,43$ for IC\,3639, NGC\,3982, NGC\,5283, NGC\,5427,
respectively) well above the minimum needed to have a BLR. In the case
of NGC\,5283, the absorption-corrected 2--10 keV luminosity that we
obtain from the X-ray spectral fit would be by itself greater than the
minimum required. Therefore none of the four is a true Seyfert 2
according to the
\citet{laor03} model.

Observations of these four objects show that it can be misleading if we
depend only on the hardness ratio to estimate absorption for sources
with a small number of counts. For example, NGC\,3982 has an HR =
$-0.7$, which is the HR expected with ACIS-S for a canonical unabsorbed
AGN spectrum. What is being measured, however, is the HR of the NLR,
with some hard-band contamination from the partially-obscured nuclear
source. This is of special concern in cases where very few counts are
detected and the hardness ratio is the only quantity that
can be determined with any certainty. If any extended soft emission is
not taken into account, the true absorption will be underestimated, and
the calculated X-ray luminosity will be too low. Even when spectral
fitting is possible, if the spectrum is not of sufficient quality to
distinguish the two components, an absorbed AGN may mimic an unobscured
AGN of lower luminosity. Several, or all, of the apparently unobscured
Seyfert 2s that have been discovered so far
\citep[e.g. ][]{pgsz01,pb02,bcc03,ghkgl03,gz03} are probably examples of
exactly this kind of mistaken identity. None of these objects has
observations with the angular resolution and high signal-to-noise
spectrum necessary to rule out the scenario we propose for our
sample. Faint AGN at moderate redshift, such as those being found in the
deep X-ray surveys, are particularly susceptible to such
misclassification. These sources are extremely important because some
intrinsically high-luminosity AGN may be masquerading as apparently
low-luminosity Seyfert 1s. These obscured AGN are exactly the sources
invoked to explain the spectrum of the cosmic X-ray background
\citep{bh05}. Also affected is the estimation of the AGN luminosity function.

\section{Summary}
\label{sec:sum}

We identified four Seyfert 2 nuclei --- IC\,3639, NGC\,3982, NGC\,5283,
and NGC\,5427 --- that appeared to be unobscured based on optical and UV
\textit{HST} data, yet did not have broad lines in either direct or polarized
light. The apparent lack of obscuration made these good ``true'' Seyfert
2 candidates, that is AGNs that genuinely lack a BLR. Since there still
could be obscuration on scales smaller than that resolvable by
\textit{HST}, we obtained X-ray observations, as X-rays are
presumed to originate close to the central engine and therefore may
``see'' absorbing material that exists on very small scales, yet hard
X-rays have high enough energy that they are expected to penetrate the
absorbing material. A lack of X-ray absorption would point to the
lack of an obscured BLR. However, all four nuclei show signs
of obscuration when observed in X-rays. Spectral fits were possible for
IC\,3639 and NGC\,5283, and imaging and spectra indicate an unresolved,
moderately absorbed ($N_H \sim 10^{21-22}$ cm$^{-2}$) nucleus with a
hard spectrum surrounded by extended soft emission. Data for the other
two objects are consistent with this picture although their X-ray
spectra are of much lower signal-to-noise ratio. A comparison of the X-ray
and optical images suggests that the soft X-ray emission arises in the
narrow-line region. Finally, we note that in cases where the extended
emission region is not resolved, and the spectrum obtained does not have
a high enough signal-to-noise ratio, it may not be to possible to
distinguish the nuclear and non-nuclear components. In such cases an
intrinsically bright but absorbed AGN may be mistaken for an
intrinsically faint, unabsorbed AGN.

\acknowledgments

Support for this work was provided by the National Aeronautics and Space
Administration through Chandra Award Number GO4-5114A issued by the Chandra
X-ray Observatory Center, which is operated by the Smithsonian
Astrophysical Observatory for and on behalf of the National Aeronautics
Space Administration under contract NAS8-03060.

%%%%%%%%%%%%%%%%%%%%%%%%%%%%%%%%%

{\it Facilities:} \facility{CXO (ACIS)}, \facility{HST (WFPC2)}

{\it Data Sets:} \dataset[ADS/Sa.CXO#Obs/4844]{4844},
\dataset[ADS/Sa.CXO#Obs/4845]{4845},
\dataset[ADS/Sa.CXO#Obs/4846]{4846}, \dataset[ADS/Sa.CXO#Obs/4847]{4847}

%%%%%%%%%%%%%%%%%%%%%%%%%%%%%%%%%

%---------------------------------------------------------------------------
%
% Tables
%

\clearpage

%
% Table 1: Chandra Observations
%

\begin{deluxetable}{lrccrcc}
\tablecaption{\textit{Chandra} Observations\label{tab:obs}}
\tablecolumns{7}
\tablehead{
\colhead{Target} & \colhead{z} & \multicolumn{2}{c}{Coordinates (J2000)} &
\colhead{Obs. Date} & \colhead{ObsID} &
\colhead{Exp. Time} \\
\colhead{} & \colhead{} & \colhead{RA} & \colhead{Dec} & \colhead{} & 
\colhead{} & \colhead{(ks)}
}
\startdata
\object{IC 3639} & 0.011 & 12 40 52.9 & $-36$ 45 22.0 & 2004 Mar 07 & 4844 & 8.7\\
\object{NGC 3982} & 0.004 & 11 56 28.1 & +55 07 31.0 & 2004 Jan 03 & 4845 & 9.2\\
\object{NGC 5283} & 0.010 & 13 41 05.7 & +67 40 20.0 & 2003 Nov 24 & 4846 & 8.9\\
\object{NGC 5427} & 0.009 & 14 03 26.1 & $-06$ 01 51.0 & 2004 Mar 26 & 4847 & 8.8\\
\enddata
\end{deluxetable}

%
% Table 2: X-Ray Measurements
%

\begin{deluxetable}{lrrrc}
\tablewidth{0pt}
\tablecaption{X-ray Measurements\label{tab:det}}
\tablecolumns{5}
\tablehead{
\colhead{Target} & \multicolumn{3}{c}{Counts} & \colhead{HR\tablenotemark{b}}\\

\colhead{} & \colhead{Broad\tablenotemark{a}} &
\colhead{Hard\tablenotemark{a}} & \colhead{Soft\tablenotemark{a}} &
\colhead{}}
\startdata
IC\,3639 & $283.4\pm 16.9$ & $24.8\pm 5.0$ & $258.7\pm 16.2$ & $-0.83\pm 0.03$ \\
NGC\,3982 & $73.0\pm 8.7$ & $11.0\pm 3.3$ & $62.1\pm 8.0$ & $-0.70\pm 0.08$ \\
NGC\,5283 & $454.4\pm 21.4$ & $337.4\pm 18.4$ & $117.1\pm 10.9$ &
$+0.48\pm 0.04$ \\
NGC\,5427 & $34.8\pm 6.0$ & $1.8\pm 1.4$ & $33.0\pm 5.8$ & $-0.90\pm 0.08$ \\
\enddata
\tablenotetext{a}{ All counts are net counts
after background subtraction. Energy bands are defined as follows: Broad 0.3--8.0 keV, Hard 2.5--8.0 keV, Soft 0.3--2.5 keV.}
\tablenotetext{b}{Hardness ratio, HR = (H$-$S)/(H+S), where H and S are
the net counts in the hard and soft bands respectively.}
\end{deluxetable}

%
% Table 3: X-Ray Spectral Fits
%

\begin{deluxetable}{lcccccccc}
\rotate
%\tablewidth{\textheight}
\tablewidth{0pt}
\tablecaption{X-Ray Spectral Fits\label{tab:spec}}
\tablecolumns{9}
\tablehead{
\colhead{} & \colhead{Galactic} & \colhead{} & 
\multicolumn{3}{c}{Spectral Fit Parameters} & \colhead{Fig.} & \colhead{$\chi_\nu^2$(dof)}\\
\colhead{Target} & \colhead{N$^{\mathrm a}_{\mathrm H}$} & 
\colhead{Model\tablenotemark{b}} & \colhead{kT(keV)} &
\colhead{N$^{\mathrm a}_{\mathrm H}$} & \colhead{$\Gamma$} & 
\colhead{} & \colhead{}
}
\startdata
IC\,3639\hspace{1em}\enspace all & 4.89 & \texttt{ab(br+pl)} & $0.11 \pm 0.03$ & $84 \pm 24$ &
$2.61 \pm 0.60$ & \ref{fig:ic3allspec} & $0.91 (7)$ \\

NGC\,3982  all & 1.23 & \texttt{ga(pl)} & \nodata & \nodata & $
3.7\pm 0.9$ & \ref{fig:n39spec} & $3.3 (2)$ \\

NGC\,5283  all & 1.84 & \texttt{ga(br+ab(pl))} & $0.71 \pm 0.27$ & $730
\pm 395$ & $0.76 \pm 0.66$ & \ref{fig:n52spec} & $0.69 (16)$\\

\phm{NGC\,5283}  extended& & \texttt{ga(br+ab(pl))} & $0.69 \pm 0.33$
& $669 \pm 28$ & $-0.13 \pm 1.16$ & \ref{fig:n52specsep} & $0.30 (5)$\\

\phm{NGC\,5283}  core & & \texttt{ga(br+ab(pl))} & $0.58 \pm 0.39$ & $985
\pm 8$ & $1.46 \pm 0.34$ & \ref{fig:n52specsep} & $0.60 (18)$\\

NGC\,5427  all & 2.38 & \texttt{ga(br)} & $0.7 \pm 0.3$ &\nodata &\nodata
& \ref{fig:n54spec} & $0.2 (4)$\\
\enddata
\tablenotetext{a}{Column density in units of $10^{20}$
cm$^{-2}$ \citep{dl90}.}
\tablenotetext{b}{Model labels: \texttt{ab}=\texttt{xswabs},
photoelectric absorption; \texttt{ga}=\texttt{xswabs} with value frozen
at Galactic column density towards this target;
\texttt{br}=\texttt{xsbremss}, thermal bremsstrahlung;
\texttt{pl}=\texttt{powlaw1d}, one-dimensional power law.}
\end{deluxetable}

%---------------------------------------------------------------------------
%
% Figures
%

\clearpage

\begin{thebibliography}{}

\bibitem[Alloin et al.(1985)]{a85} Alloin, D., Pelat, D., Phillips, M.,
\& Whittle, M. 1985, \apj, 288, 205

\bibitem[Antonucci(1993)]{ski93} Antonucci, R. 1993, \araa, 31, 473

\bibitem[Barcons, Carrera, \& Ceballos(2003)]{bcc03} Barcons, X.,
Carrera, F. J., \& Ceballos, M. T. 2003, \mnras, 339, 757

\bibitem[Bianchi, Guainazzi \& Chiaberge(2006)]{bgc05} Bianchi, S.,
Guainazzi, M., \& Chiaberge, M. 2006, \aap, 448, 499

\bibitem[Boisson \& Durret(1986)]{bd86} Boisson, C., \& Durret, F. 1986,
in New Insight in Astrophysics, 8 Years of UV Astronomy with IUE
(ESA SP-263), 687, ed. Rolfe, E.J.

\bibitem[Brandt \& Hasinger(2005)]{bh05} Brandt, W. N., \& Hasinger,
G. 2005, \araa, 43, 827

\bibitem[Cohen et al.(1986)]{c86} Cohen, R.D., Rudy, R.J., Puetter, R.C.,
Ake, T.B., \& Foltz, C.B. 1986, \apj, 311, 135

\bibitem[Dickey \& Lockman(1990)]{dl90} Dickey, J. M., \& Lockman,
F. J. 1990, \araa, 28, 215

\bibitem[Dopita et al.(2002)]{dpkc02} Dopita, M. A., Pereira, M.,
Kewley, L. J., \& Capaccioli, M. 2002, \apjs, 143, 47

\bibitem[Garcia-Rissmann et al.(2005)]{grea05} Garcia-Rissmann, A.,
Vega, L. R., Asari, N. V., Cid Fernandes, R., Schmitt, H., Gonz\'{a}lez
Delgado, R. M., \& Storchi-Bergmann, T. 2005, \mnras, 359, 765

\bibitem[Georgantopoulos \& Zezas(2003)]{gz03} Georgantopoulos, I., \&
Zezas, A. 2003, \apj, 594, 704

\bibitem[Gezari et al.(2003)]{ghkgl03} Gezari, S., Halpern, J. P.,
Komossa, S., Grupe, D., \& Leighly, K. M. 2003, \apj, 592, 42

\bibitem[Guainazzi et al.(2001)]{g01} Guainazzi, M., Fiore, F., Matt, G.,
\& Perola, G.C. 2001, \mnras, 327, 323

\bibitem[Heckman et al.(2004)]{hea04} Heckman, T. M., Kauffmann, G.,
Brinchmann, J., Charlot, S., Tremonti, C., \& White, S. D. M. 2004,
\apj, 613, 109

\bibitem[Heckman et al.(1995)]{h95} Heckman, T., et al. 1995, \apj, 452, 549

\bibitem[Heckman et al.(2005)]{hphk05} Heckman, T. M., et al. 2005,
\apj, 634, 161

\bibitem[Hornschemeier et al.(2005)]{hhptc05} Hornschemeier, A. E., et
al. 2005, \aj, 129, 86

\bibitem[Huchra \& Burg(1992)]{hb92} Huchra, J., \& Burg, R. 1992,
\apj, 393, 90

\bibitem[Kennicutt(1998)]{k98} Kennicutt, R. C. 1998, \apj, 498, 541

\bibitem[Khachikian \& Weedman(1974)]{kh74} Khachikian, E.Y., \&
Weedman, D.W. 1974, \apj, 192, 581

\bibitem[Laor(2003)]{laor03} Laor, A. 2003, \apj, 590, 86

\bibitem[Levenson et al.(2006)]{lhkwz06} Levenson, N. A., Heckman,
T. M., Krolik, J. H., Weaver, K. A., \& \.{Z}ycki, P. T. 2006, \apj, in
press (astro-ph/0605438)

\bibitem[Lumsden, Alexander, \& Hough(2004)]{lah04} Lumsden, S.L.,
Alexander, D.M., \& Hough, J.H. 2004, \mnras, 348, 1451

\bibitem[Lumsden et al.(2001)]{l01} Lumsden, S.L., Heisler, C.A., Bailey, 
J.A., Hough, J.H., \& Young, S. 2001, \mnras, 327, 459

\bibitem[Maiolino et al.(1998)]{maiolino98} Maiolino, R., Salvati, M.,
Bassani, L., Dadina, M., della Ceca, R., Matt, G., 
Risaliti, G., \& Zamorani, G. 1998, \aap, 338, 781

\bibitem[Malkan, Gorjian, \& Tam(1998)]{mgt98} Malkan, M.A., Gorjian, V.,
\& Tam, R. 1998, \apjs, 117, 25

\bibitem[Moran et al.(2000)]{moran00} Moran, E.C., Barth, A.J.,
Kay, L.E., \& Filippenko, A.V. 2000, \apj, 540, L73

\bibitem[Mulchaey et al.(1994)]{mea94} Mulchaey, J. S., Koratkar, A.,
Ward, M. J., Wilson, A. S., Whittle, M., Antonucci, R. R. J., Kinney,
A. L., \& Todd, H. 1994, \apj, 436, 586

\bibitem[Murray \& Chiang(1998)]{mc98} Murray, N., \& Chiang, J. 
1998, \apj, 494, 125

\bibitem[Nelson \& Whittle(1995)]{nw95} Nelson, C. H., \& Whittle, M.
1995, \apjs, 99, 67

\bibitem[Nicastro(2000)]{n00} Nicastro, F. 2000, \apj, 530, L65

\bibitem[Nicastro et al.(2003)]{n03} Nicastro, F., Martocchia, A., \&
Matt, G. 2003, \apj, L13

\bibitem[Osterbrock \& Martel(1993)]{om93} Osterbrock, D. E., \& Martel,
A. 1993, \apj, 414, 552

\bibitem[Panessa \& Bassani(2002)]{pb02} Panessa, F., \& Bassani,
L. 2002, \aap, 394, 435

\bibitem[Pappa et al.(2001)]{pgsz01} Pappa, A., Georgantopoulos, I.,
Stewart, G. C., \& Zezas, A. L. 2001, \mnras, 326, 995

\bibitem[Penston \& Perez(1984)]{pp84} Penston, M.V., \& Perez, E.
1984, \mnras, 211, 33

\bibitem[P\'{e}rez Garc\'{i}a \& Rodr\'{i}guez Espinosa(2001)]{pgre01}
P\'{e}rez Garc\'{i}a, A. M., \& Rodr\'{i}guez Espinosa, J. M. 2001,
\apj, 557, 39

\bibitem[Peterson(1997)]{peterson97} Peterson, B.M. 1997, An Introduction
to Active Galactic Nuclei, Cambridge Univ. Press

\bibitem[Peterson et al.(2004)]{peterson04} Peterson, B.M., et al. 2004,
\apj, 613, 682

\bibitem[Pogge \& Martini(2002)]{pm02} Pogge, R.W., \& Martini, P. 2002, 
\apj, 369, 624

\bibitem[Pogge et al.(2006)]{pip06} Pogge, R. W., et al. 2006, in prep.

\bibitem[Regan \& Mulchaey(1999)]{rm99} Regan, M.W., \& Mulchaey, J.S.
1999, \aj, 117, 2676

\bibitem[Sandage \& Tammann(1987)]{st87} Sandage, A., \& Tammann, G.A.
1987, A Revised Shapley-Ames Catalog of Bright Galaxies (Publ. 635; 
Washington D.C.; Carnegie Institute of Washington)

\bibitem[Tran(2001)]{tran01} Tran, H.D. 2001, \apj, 554, L19

\bibitem[Whittle(1992)]{w92a} Whittle, M. 1992, \apjs, 79, 49
\end{thebibliography}
\end{document}